\documentclass[manuscript]{aastex}

\usepackage{multirow}

\shorttitle{thermal structure of coronal holes}
\shortauthors{Hiremath}

\begin{document}


\title{THERMAL AND MAGNETIC FIELD STRUCTURE OF NEAR EQUATORIAL CORONAL HOLES}


\author{Hiremath, K. M. and Hegde, M}
\affil{Indian Institute of Astrophysics, Bangalore, India}
\email{hiremath@iiap.res.in}

\begin{abstract}
{We use full-disk, SOHO/EIT  195 $\AA$  calibrated  images to measure 
latitudinal and day to day variations of area and average photon 
fluxes  of the near equatorial coronal holes. In addition, 
energy emitted by the 
coronal holes with their temperature and strength of magnetic 
field structures are estimated. By analyzing data of 2001-2008, we find that variations of average area (A), 
photon flux (F), radiative energy (E)  and temperature (T) of coronal holes
 are independent of latitude. Whereas inferred strength of magnetic field structure of
 the coronal holes is dependent on the latitudes and varies
from low near the equator to high near both the poles. 
 Typical average values of estimated physical parameters are: 
$A \sim 3.8(\pm0.5)\times10^{20}~cm^{2}, F \sim 2.3(\pm0.2)\times10^{13}~photons\;cm^{-2}\;sec^{-1}, 
E \sim 2.32(\pm0.5)\times 10^{3}~ergscm^{-2}sec^{-1} \ and \ T \sim 0.94(\pm0.1)\times10^{6} ~$ K.
Average strength of magnetic field structure of coronal hole at the
corona is estimated to be $\sim$ $0.08 \pm 0.02$ Gauss.
If coronal holes are anchored in the convection zone, one would expect they
 should rotate differentially. Hence, thermal wind balance and 
isorotation of coronal holes with the solar plasma implies the temperature 
difference between
the equator and both the poles. Contrary to this fact, variation of 
thermal structure of near equatorial coronal holes is independent
of latitude leading to a conclusion that coronal holes must rotate 
rigidly that are likely to be anchored initially below the tachocline confirming our previous study (ApJ, 763, 137, 2013).}
\end{abstract}

\keywords{sun:equatorial coronal holes, sun:thermal structure, Coronal Hole:thermal structure, 
Coronal Hole:magnetic field structure}

\section{INTRODUCTION}
Coronal holes (CH), with unipolar magnetic field structures (Harvey \& Sheeley 1979; Harvey et al. 1982),
 are the lowest density plasma structures mainly detected in
either UV or X-ray radiations of the sun's atmosphere and are associated with
rapidly expanding magnetic fields and the acceleration of the high-speed solar 
wind (Krieger et al. 1973; Neupert \& Pizzo, 1974; Nolte et al. 1976; Zirker 1977; 
Cranmer 2009 and references therein; Wang 2009; Wiegelmann, Thalmann, 
Solanki 2014). Possible link of sunspot
(Hiremath and Mandi 2004 and references there in; Hiremath 2009)
activity on the Earth's atmosphere and climate are well recorded
in the literature. Recently, evidences
are building up that coronal holes also trigger responses in the Earth's upper atmosphere 
and magnetosphere (Soon et al. 2000; Lei et al. 2008;
Shugai et al. 2009; Sojka et al. 2009;
Choi et al. 2009;  Ram, Liu \& Su 2010; Krista 2011; 
Verbanac et al. 2011; Fathy {\em et.al.} 2014; Machiya and Akasofu 2014).
Very recently, Hiremath, Hegde and Soon (2015) came to a conclusion that, in addition
to influence of sunspots, emission of coronal holes also 
trigger and maintain Indian Monsoon rainfall. 

From the ISRO (Indian Space Research Organization)
 funded project, authors$'$ interest in the study of coronal holes is to examine
radiative responses of these structures on the Indian summer
Monsoon rainfall. For this purpose, estimation of radiative flux and energy of
the CH at 1 AU during their evolution passage on the solar disk is necessary.
Keeping these main objectives in mind, by using SOHO/EIT  195 $\AA$  calibrated  images,
thermal structure of CH such as photon flux, energy and hence temperature at the
sun and at 1 AU are estimated.

Dynamics such as rotation rates (Hiremath and Hegde 2013 and references there in) and 
{\em in situ} plasma conditions of the CH are also of considerable interest for the solar
community as the fast solar wind  most likely originates in these regions (Stucki et al. 
2002; Hegde {\em et.al.} 2015). 
Ion temperatures in a polar coronal hole from the line width measurements are 
estimated by Tu et al. (1998). 
Centre-to-limb variation of radiance of the transition region and coronal lines have been obtained by Wilhelm et al. (1998). 
Doppler shifts measurements of CII, O VI and Ne VIII lines are obtained by 
Warren, Mariska and Wilhelm(1997). Peter \& Judge (1999) also studied these and few other lines and found that the redshift to blue shift transition occurs 
at electron temperatures of about 5 $\times10^{5}$ K.
Analysis of SXT/Yohkoh images (Hara et. al. 1996) have shown that the estimated temperature 
structure (1.8 -2.4  $\times10^{6}$ K) 
 in CH is of same order as that in the ambient medium. Mogilevsky, Obridko and Shilova (1997) also arrive at a similar conclusion. From the analysis of Big Bear solar observatory 
magnetograms and SOHO EIT images, Zhang {\em et.al.} (2007) came to a
conclusion that temperature of coronal hole and ambient medium is not entirely
different. 

 In recent years, space based observations such as  
Yohkoh and SOHO missions are extensively used for
estimation of temperature structure of CH (Hara et al. 1994; Moses et al. 1997).
One physical parameter that senses thermal structure of CH is electron 
temperature that is estimated
by different spectroscopic methods. A detailed assessment of observations in coronal holes 
and the deduced temperatures is published by Habbal, Esser and Arndt (1993). 
See also
a detailed review of estimation of coronal hole
temperature by Habbal (1996) and Wilhelm (2012) respectively. Electron 
temperatures in CH can be measured with the help of a Magnesium line 
ratio of a temperature-sensitive pair (see Wilhelm 2006). With the assumption that
density and temperature of the gas from which spectral lines are formed are constant along 
the line of sight, Habbal et al. (1993) estimated the temperature structure of 
CH.
Using two SOHO spectrometers, CDS and SUMER, electron temperature of CH is 
measured as a function of height above the limb
in a polar coronal hole (David et al. 1998; Wilhelm et al. 1998). Doschek \& Laming 
(2000) found that  increase in the emission-line ratio of the polar coronal 
hole is primarily due to  
increase of the electron temperature with height. Marsch, Tu and Wilhelm  (2000) found that 
the hydrogen temperature increased only slightly from 1 $\times 10^{5}$ to 
2 $\times10^{5}$ K in the height range 
from 12,000 to 18,000 km. Stucki et al. (2000) presented that with increasing formation
temperature, spectral lines show on average, an increasingly stronger blue shift 
in coronal holes relative to the quiet sun at equal heliospheric angle. 
Furthermore, Xia, Marsch and Wilhelm  (2004) reported that the bases of coronal holes
seen in chromospheric spectral lines with relatively low formation temperatures 
displayed similar properties as normal
quiet sun regions. More recently, studies of Wilhelm (2006; 2012) 
suggest that, in a polar 
coronal hole region, the electron temperature in plumes is estimated to be 
$\sim 8 \times10^{5}$ K and, $\sim 1.13 \times 10^6$ 
K in the inter plume regions, at the height of 45 Mm above the limb.

From the silicon and iron coronal lines, Doscheck and Feldman (1977)
conclude that CH temperature must be less than 1 MK. 
From the jets of coronal holes, Nistico {\em et.al.} (2011) compute electron
temperature of CH by the filter ratio method at 171 $\AA$ and 195 $\AA$ and estimated
the temperature structure with a magnitude in the range of $\sim$ 0.18-1.3 MK.
By using two lines Mg IX (368 \AA) and Mg X (625 \AA) and, with a similar 
line ratio technique, Doyle {\em et.al.} (2010) derived the different 
coronal hole temperature structures during solar maximum ($\sim$ 1.04 MK)
and during minimum ($\sim$ 0.82 MK) respectively.
The EUV (Fe XV at 284 \AA) and radio (at 169 and 408 MHz) observations
(Chiuderi, Avignon and Thomas 1977) suggest that CH consists of
a hot (10 \% of the CH surface with $\sim$ $2 \times 10^{6}$ K)
and a cold ($\sim$ $0.8 \times 10^{6}$ K) regions to explain observations of 
both the EUV and radio coronal hole 
temperature structures. From the
analysis of EUV (SOHO/CDS) and  radio emission (164-410 MHz, 
by the Nancay Radioheliograph, France), and with a model, 
Chiuderi {\em et. al.} (1999) estimated the CH temperature
structure to be $\sim$ $9 \times 10^{5}$ K.
From the spectroscopic diagnostics of Mg VIII (430.47 $\AA$ and 436.62 \AA) ion,
observed by Sky lab space probe, Dwivedi and Mohan (1995), estimated the
coronal hole electron temperature to be $\sim$ $8 \times 10^{5}$ K.
Dwivedi, Mohan and Wilhelm (2000), from the SUMER observations,
estimate the CH temperature to be $\sim$ $(6.5-7.5) \times 10^{5}$ K.
Observations (Esser {\em et.al.} 1999) of Ly$\alpha$, 
$\lambda$1216, Mg X (625 \AA), and O VI (1038 \AA) 
spectral lines with the UVCS instrument on board SOHO, 
from $1.35-2.1 R_{\odot}$, yield that
proton temperature of CH slowly increases between 1.35 and $2.7 R_{\odot}$
 and does not exceed $3 \times 10^{6}$ K in that region.

In most of the afore mentioned studies, estimated electron temperature, along
the slit of the observations over the region of CH, does not represents temperature 
structure of whole CH. Using filter ratio technique, 
although one can use DEM method for
estimation of temperature of CH, this method yields ambiguity (Zhang, White and Kundu,  1999; Landi, Reale \& Testa 2012; Guennou {\em et al.} 2012). 
Moreover, line
ratio method might not be appropriate ( see section 4.5, 
Slemzin, Goryaev and Kuzin 2014) for estimation of coronal hole temperature
  as there is a appreciable contribution of scattered light due to 
brighter ambient corona, especially for the coronal holes. However, in this study we present a simple method for estimation of radiative flux, energy
and temperature structure of CH at the sun and near Earth. 

As solar wind due to coronal hole is directly proportional to area 
(Hegde {em et.al.} 2015; Rotter {\em et.al} 2012; 
Karachik and Pevtsov 2011; Abramenko {\em et. al.} 2009; 
Shugai, Veselovsky and Trichtchenko 2009; Vršnak, Temmer and Veronig 2007;
Nolte {\em et.al.} 1976), it is interesting to examine how these 
parameters vary during
the evolution of coronal hole. To be specific, for example, as the Earth's
ionosphere responds with the solar wind due to coronal hole it is
important to estimate physical parameters such as area, temperature
structure, etc., 

As for dynamics, except some of the studies (Shelke \& Pande 1985; 
Obridko \& Shelting 1989; Navarro-Peralta
\& Sanchez-Ibarra 1994; Insley et al. 1995), other studies
(Wagner 1975; Wagner 1976; Timothy \& Krieger 1975; Bohlin 1977;
Hiremath and Hegde 2013; Japaridze {\em et.al.} 2015) 
indicate rigid body rotation rates of the coronal holes. 
With large number of data and accurately estimated
average longitude from the central meridian of the
coronal holes, especially Hiremath and Hegde (2013) came to
a conclusion that irrespective of area and latitudes coronal holes
rotate rigidly.
 We do not mean that all
the coronal holes rotate rigidly. This is because all means, even the
coronal holes near the poles. However, we have restricted
the data (also with additional three criteria as described in
section 2) of coronal holes that occur between 40 deg North to 40 deg
South, that is near equatorial coronal holes. 

As the coronal
holes are unipolar magnetic flux tubes (Harvey \& Sheeley 1979; 
Harvey et al. 1982), condition
of infinite conductivity of the corona leads to isorotation
of the coronal hole flux tubes with the ambient plasma
rotation. If the coronal holes rotate differentially, then
thermal wind balance equation (Brun, Antia and Chitre 2010)
 yields temperature difference between
the equator and the poles. On the other hand, if coronal holes 
rotate rigidly, there is no temperature difference between
the equator and poles. Hence, in order to confirm whether
coronal holes rotate rigidly or differentially, information
regarding latitudinal variation of thermal structure of the 
coronal holes is necessary.

Physics of MHD waves (that emanate from the coronal
holes) is important not only for understanding the
heating of corona but also useful for understanding
the fast solar wind. Many MHD models (Davila 1985; 
Cally 1986, 1987; Ofman 2005 and references there in) 
were developed to probe these phenomena where in
strength and geometry of magnetic field structure
of coronal hole are necessary.  Although a general consensus
is emerged that geometry of coronal hole magnetic
field structure is unipolar, to the knowledge of
authors and till date, no study is available that estimates magnetic field
strength of coronal hole in 195 $\AA$ that
probably originates (Fig 2, Yang {\em et. al.} 2009) 
around 1.1 solar radius.

In addition to importance of physical
parameters of CH for study of solar-terrestrial relationship, 
one has to also address the following 
questions in order to resolve the fundamental
problems presented in the previous paragraphs. To summarize
the same: 
(i) How the physical properties, such as area, 
radiative flux, energy and temperature
structure vary during the evolution passage of CH over the observed solar disk?
(ii) Do these physical parameters of CH are dependent or independent
of heliographic latitude? (iii) From the information of latitudinal
variation of temperature structure of CH, is it possible to
get any information whether CH rotate rigidly or differentially?
(iv) What is strength of magnetic field structure of CH at the height of coronal region around 1.1$R_{\odot}$ where the line 195 $\AA$ originates.

In order to seek the answers to afore mentioned problems,
we use near equatorial coronal holes for the the present study.
 
In section 2, we present the data of near equatorial
coronal holes and method of analysis, 
and the results of this analysis are presented in section 3. In section 4, 
with a brief discussion, 
conclusions of this study are presented.

\section{DATA AND ESTIMATION OF DIFFERENT PHYSICAL PARAMETERS OF CH}
We use 195 $\AA$ full-disk images obtained by EIT on board SOHO, although 
on board instrument also observes full-disk EUV images in 
other wave length (171 $\AA$, 284 $\AA$ and 304 $\AA$) pass bands. 
A detailed description of the instrument is provided by Delaboudini\'{e}re et al. (1995).
The obtained images are in FITS format and individual pixels are in units of
data number (DN/sec). DN is defined to be
output of the instrument electronics  which
corresponds to the incident photon signal  converted into charge within
each CCD pixel (Madjarska \& Wiegelmann 2009).
Further details of SOHO/EIT 195 $\AA$ images, their calibration,
method of detection of CH with estimation of heliographic
coordinates (such as latitude $\theta$ and longitude $l$ ) and, 
computation of total DN counts (TDN) of coronal holes
are described in our previous study (Hiremath \& Hegde 2013). 
 In the present study, we mainly concentrate on the 
 data of near equatorial coronal holes that are distributed
with in 40 deg north to 40 deg south. Additional three more criteria
 used in selection of the data are: 
(i) in order to minimize the projection effects (especially
coronal holes near both the eastern and the western limbs), we considered only
the coronal holes that emerge within 65$^\circ$  central meridian distance,
(ii) the coronal holes must be compact, independent, and not 
elongated in latitude and,  (iii) during coronal holes 
passage across the solar disk, it should not merge with other coronal hole.

\subsection{Computation of Area of CH}
As described in the previous study (Hiremath and Hegde 2013), 
once boundary of a CH
is detected, total number of pixels (TNP) with in the detected
boundary is estimated and
area $A$ of coronal hole and its measured uncertainty $\delta A$ are
computed as follows. 
\begin{equation}
\displaystyle A = c_{1} \frac{TNP}{cos\;l} \hspace{0.7cm} cm^{2} \,, 
\end{equation}
\begin{equation}
 \displaystyle\delta A = c_{1} (TNP) (tan\;l sec\;l) \delta l \hspace{0.7cm} cm^{2} \, ,
\end{equation}
where the multiplicative constant term $c_{1} (=3.573\times10^{16}$)
is estimated from the resolution of pixel size and 
 the factor $1/(cos\;l)$ ($l$ is viewing angle or longitude from 
the central meridian) is a correction factor for the
projectional effect for the CH that are close to the limb. 

We have also corrected the projectional
effects from the formula $A_{c}=c_{1} A_{obs}/cos(\delta)$ (where $A_{c}$
is corrected area, $A_{obs}$ is observed area and
$cos(\delta)=sin(B_{0})sin(\theta)+cos(B_{0})cos(\theta )cos(l)$,
with $\theta$ and $l$ are heliographic latitude and heliographic 
longitude from the central meridian of the CH respectively,
whereas $B_{0}$ is the heliographic latitude of the center of the 
solar disk at the time of observation) that takes into account both
the latitude and longitudinal projections. However,
we got the same results. This is obvious as the data set is
not in the higher latitudes.

\subsection{Computation of Average Radiative Flux of CH at $L_{1}$}
For estimation of radiative flux $F$ of CH at the Lagrangian
point $L_{1}$ in the space, we use information from the SOHO/EIT instrumental
response curve (see the Figure 3, a postscript
file $calib.ps$ is obtained from the website http://umbra.nascom.nasa.gov/eit/eit\_guide/). 
However, according to SOHO/EIT website information, CH data in 195 $\AA$ is also
sensitive to other three wavelength (171 $\AA$, 284 $\AA$ and 304 $\AA$) bands
whose contributions to the instrumental responses are to be computed
judiciously in the following way. 
For this purpose, by integration of area under curve (Fig 3),
 one has to estimate response values of $R_{1}$, $R_{2}$, $R_{3}$
and  $R_{4}$ for all the four (171 $\AA$, 195 $\AA$, 284 $\AA$ and 304 $\AA$) wavelength channels.
First we manually digitize all the four response curves (see Table 1)
and by using Trapezoidal rule method, integration of area under curve is computed.
Finally a grand average response R (=($R_{1}$+$R_{2}$+$R_{3}$+$R_{4}$)/4)
is computed. Results of average responses for different channels
are presented in Table 1.

 \begin{table}
  \centering
      \caption{Digitized values of the instrumental response curves}
  \begin{tabular}{@{}lllll@{}}
  \hline
$\lambda$ & R1 & R2 &  R3 &  R4 \\
$\AA$ & & & &\\
\hline
170 & 2e-15   & 1.0e-16 &1.0e-16 & 1.0e-16 \\
180 & 1.5e-12 & 3e-13   &7e-16   & 5e-15\\
190 & 5e-14 & 6e-12     &8e-16   & 4e-15\\
200 & 1.5e-14 & 8e-13   &9e-16   & 4e-15\\
210 & 6e-15 & 1.5e-13   &1e-15   & 5e-15\\
220 & 8e-15 & 7e-14     &1.5e-15 &6e-15\\
230 & 6e-15 & 5e-14     &1.5e-15 &7e-15\\
240 & 3e-15 & 3e-14     &2.5e-15 &9e-15\\
250 & 2.5e-15 & 2e-14   &4e-15   &1.1e-14\\
260 & 1.5e-15 & 1.5e-14 &2e-14   &1.5e-14\\
270 & 1.5e-15 & 9e-15   &1.1e-13 &3e-14\\
280 & 9e-16 & 7e-15     &3e-13   &7e-14\\
290 & 6e-16 & 4e-15     &2e-14   &1.5e-13\\
300 & 4e-16 & 2e-15     &3e-16   &4e-13\\
310 & 1.5e-16 & 9e-16   &        &3.8e-13\\
320 &         & 4e-16   &        &9e-14\\
330 &         & 2e-16   &        & 9e-15\\
340 & & & &1.2e-15\\
\hline
avg=&4.43465e-15 & 1.75493e-14&1.40760e-15&2.60490e-15\\
 \end{tabular}
\tablenotetext{}{Note: Unit of instrumental responses ($R_{1}\, to \, R_{4}$) is (DN
$sec^{-1}$)/(photons $cm^{-2}$ $sec^{-1}$ $steradian^{-1}$ $\AA^{-1}$)}
 \end{table}

As the EIT instrumental response function $R$ is in the unit of DN 
$sec^{-1}$/(photons $cm^{-2}$ $sec^{-1}$ $steradian^{-1}$ $\AA^{-1}$),
one can divide measured TDN (total number of DN counts of CH) by the instrumental response function $R$ in order to get the radiative
flux emitted by CH. Hence, total radiative flux $F$ emitted by
whole region of CH is
\begin{equation}
\displaystyle F= c_{2} \frac{TDN}{R} sin \theta \, photons \, cm^{-2} sec^{-1} Sr^{-1} \, ,
\end{equation}
 and its error $\delta$ F is

$$
\displaystyle\delta F =c_{2} {\Big [} \left(\frac{TDN}{R}\right)cos \theta \delta \theta +sin\theta * \delta \left(\frac{TDN}{R}\right)\Big ] \, ,
$$

\noindent where the multiplicative constant factor $c_{2}=2.38\times10^{-6}$ is computed from
average of all the four wavelengths (171$\AA$, 195$\AA$, 284$\AA$ and 304 $\AA$) in order
to eliminate the term $\AA^{-1}$ in the instrumental response $R$.
Similarly equation of radiative flux F1 at the Lagrangian point L1 (near earth)
is 
  \begin{equation}
\displaystyle F1= c_{2} \frac{TDN}{R} sin \theta  Sr  \, photons \, cm^{-2} sec^{-1}, 
\end{equation} where

\noindent $Sr={A \over D^{2}}$ ($A$ is area of CH, Sr is steradian angle 
and $D$ is distance between the sun and the orbit of SOHO satellite).    
Uncertainty
in the radiative flux $\delta F1$ of CH is computed as follows
$$
\displaystyle\delta F1 =c_{2} {\Big [} \left(\frac{TDN}{R}\right)cos \theta Sr \delta \theta 
+\left(\frac{TDN}{R}\right) sin\theta Sr\_err 
$$

\begin{equation}
+\delta (\frac{TDN}{R})sin \theta  Sr {\Big ]} \, photons \, cm^{-2} sec^{-1}  Sr^{-1}\, ,
\end{equation}   
where Sr\_err is error in steradian.

\subsection{Computation of Average Radiative Energy emitted by CH at $L_{1}$}
Total radiative energy $E$ emitted by CH is
\begin{equation}
\displaystyle E=h\nu F_{1} \, ergs  cm^{-2} sec^{-1} \, ,
\end{equation}
where $h\nu$ ($h$ is Planck's constant and $\nu$ is frequency of radiation) 
is a quanta of photon energy of the EUV radiation. Uncertainty $\delta E$ in
the energy is 
\begin{equation}
\displaystyle\delta E = h\nu\delta F1 \, ergs  cm^{-2} sec^{-1} . 
\end{equation}

\subsection{Computation of Average Temperature structure of CH in the Corona}

Spectroscopic methods yield temperature along the observed slit. Whereas here
we measure the average temperature of the observed whole coronal hole
which we call as ``temperature structure".

From the information of radiative energy ($I_{CHL}$) of CH at the Lagrangian
point $L_{1}$, following ratio yields the radiative energy ($I_{CHS}$) of CH
in the corona
\begin{equation}
\frac{I_{CHS}}{I_{CHL}} = \frac{\int (E cos \theta d \theta) A/{R_{CH}}^{2}}
{\int (E cos \theta d \theta) A/{R_{CHL}}^{2}} = \frac{{R_{CHL}}^{2}}{{R_{CH}}^{2}} \, ,
\end{equation}
\noindent where $R_{CHL}$ is the distance between sun's center and the
Lagrangian point $L_{1}$ and, $R_{CH}$ is distance between the centre
of the sun and the height at which CH is formed in the corona.
By knowing the values of $R_{CH}$, as observed CH in 195 $\AA$ is formed at the height of $\sim$ 1.1 $R_{\odot}$
(where $R_{\odot}$ is radius of the sun; see Fig 2 of Yang {\em et.al.} 2009), and $R_{CHL}$, the ratio of RHS of
equation (8) is estimated to be $\sim$ $3.14 \times 10^{4}$. Hence, we get $I_{CHS} = 3.14 \times 10^{4} I_{CHL} 
= 3.14 \times 10^{4} (\frac{TDN}{R} sin \theta) $. 
\noindent Assuming that plasma of CH is in thermodynamic equilibrium, total energy radiated by
CH is equated with the Planck's law and average temperature structure $T$ and
 its uncertainty $\delta T$ of CH are computed as follows 
\begin{equation}
\displaystyle T = \frac{hc}{\lambda k ln(\frac {2hc^2}{\lambda^5 I_{CHS}}+1)} \, K, 
\end{equation}

\begin{equation}
\displaystyle\delta T = \left(\frac{\lambda k}{hc \; 
ln(\frac {2hc^2}{\lambda^5 I_{CHS}}+1)}\right) \frac{2hc^2 (\delta I_{CHS}) T^2}{\lambda^5 I_{CHS}^2} \, K
\end{equation}
\noindent where $c$ is velocity of light, $\lambda$ is wavelength and $k$
is Boltzmann constant respectively and, $I_{CHS}$ is radiative energy of CH 
at the corona.

\begin{table}
\begin{minipage}{250mm}
\caption{\bf Daily variation of different physical parameters of Coronal Holes }
 \begin{tabular}{@{}llllllllllll@{}}
 \hline 
\multicolumn{12}{|c|}{01-05 Jan 2001} \\
\hline
 Area  & $\delta$Area & FS$^{1}$ & $\delta$FS$^{2}$ & FAU$^{3}$ & $\delta$FAU$^{4}$ & ES$^{5}$ & $\delta$ES$^{6}$ &EAU$^{7}$ &
$\delta$EAU$^{8}$ & T$^{9}$ & $\delta$T$^{10}$ \\
E+20   & E+20          & E+13& E+13       &E+8 &  E+8         &E+3 & E+3        
&E-2 &E-2 &E+6& E+6   \\
\hline
5.124 &0.680  & 5.637  & 0.445 &1.319 &0.115  &5.694  &0.470 &1.332 &0.118   & 1.075
 &0.697 \\
5.286 &0.391  & 5.868  & 0.515 &1.416 &0.084  &5.927  &0.312 &1.431 &0.112   & 1.057
 &0.605\\
5.058 &0.199  & 5.354  & 0.514 &1.237 &0.152  &5.407  &0.341 &1.249 &157   &1.045 
&0.614\\
4.788 &0.122  &4.571  &0.215  &0.999 &0.047  &4.617  &0.312 &1.009 &0.107   &1.034  
&0.601\\
4.732 &0.101  &5.658  &0.238  &1.223 &0.561  &5.715  &0.390 &1.235 &0.523   &1.048  
&0.615\\
\hline
\multicolumn{12}{|c|}{01-06 Jan 2001} \\
\hline
Area  & $\delta$Area & FS & $\delta$FS & FAU & $\delta$FAU & ES & $\delta$ES &EAU &
$\delta$EAU & T & $\delta$T \\
E+20   & E+20          & E+12& E+12       &E+7 &  E+7         &E+2 & E+2       
&E-3 &E-3 &E+6& E+6   \\
\hline
2.790  &0.106  &1.537 &0.403 &1.958 &0.521  &1.651  &0.155 &1.978 &0.526   & 1.191 
&0.167\\
3.423 &0.060  &2.273 &0.596  &3.553 &0.936  &2.967  &0.230 &3.588 &0.945  & 1.162 
&0.222\\
3.145  &0.025  &2.242 &0.532  &3.220 &0.766  &2.430  &0.227 &3.252 &0.774   & 1.164 
&0.325\\
2.754 &0.079  &2.016 &0.411  &2.535 &0.524  &1.661  &0.204 &2.561 &0.053   & 1.153 
&0.342\\
1.642 &0.092  &2.018 &0.515  &0.861 &0.013  &4.058  &0.116 &0.870 &0.019   & 0.960 
&0.412\\
1.552 &0.177  &2.283 &0.514  &0.808 &0.012  &2.406  &0.115 &0.817 &0.077   & 0.955 
&0.356\\
\hline

\end{tabular}
\tablenotetext{1}{Average radiative flux (photons cm$^{-2} sec^{-1} Sr^{-1}$) of CH measured on the Sun}
\tablenotetext{2}{Uncertainty in radiative flux of CH measured on the Sun}
\tablenotetext{3}{Average radiative flux (photons cm$^{-2} sec^{-1}$) of CH estimated at 1AU}
\tablenotetext{4}{Uncertainty in radiative flux of CH measured at 1AU}
\tablenotetext{5}{Average radiative energy (ergs  $cm^{-2} sec^{-1}Sr^{-1}$) of CH measured on the Sun}
\tablenotetext{6}{Uncertainty in radiative energy of CH measured on the Sun}
\tablenotetext{7}{Average radiative energy (ergs $cm^{-2} sec^{-1}$) of CH estimated at 1AU}
\tablenotetext{8}{Uncertainty in radiative energy of CH measured at 1AU}
\tablenotetext{9}{Average temperature (K) of CH measured on Sun}
\tablenotetext{10}{Uncertainty in temperature of CH measured on Sun}
\end{minipage}
\end{table}



\begin{figure}
\begin{center}
    Fig 1(a) \hskip 40ex  Fig 1(b)
    \begin{tabular}{cc}
      {\includegraphics[width=18pc,height=18pc]{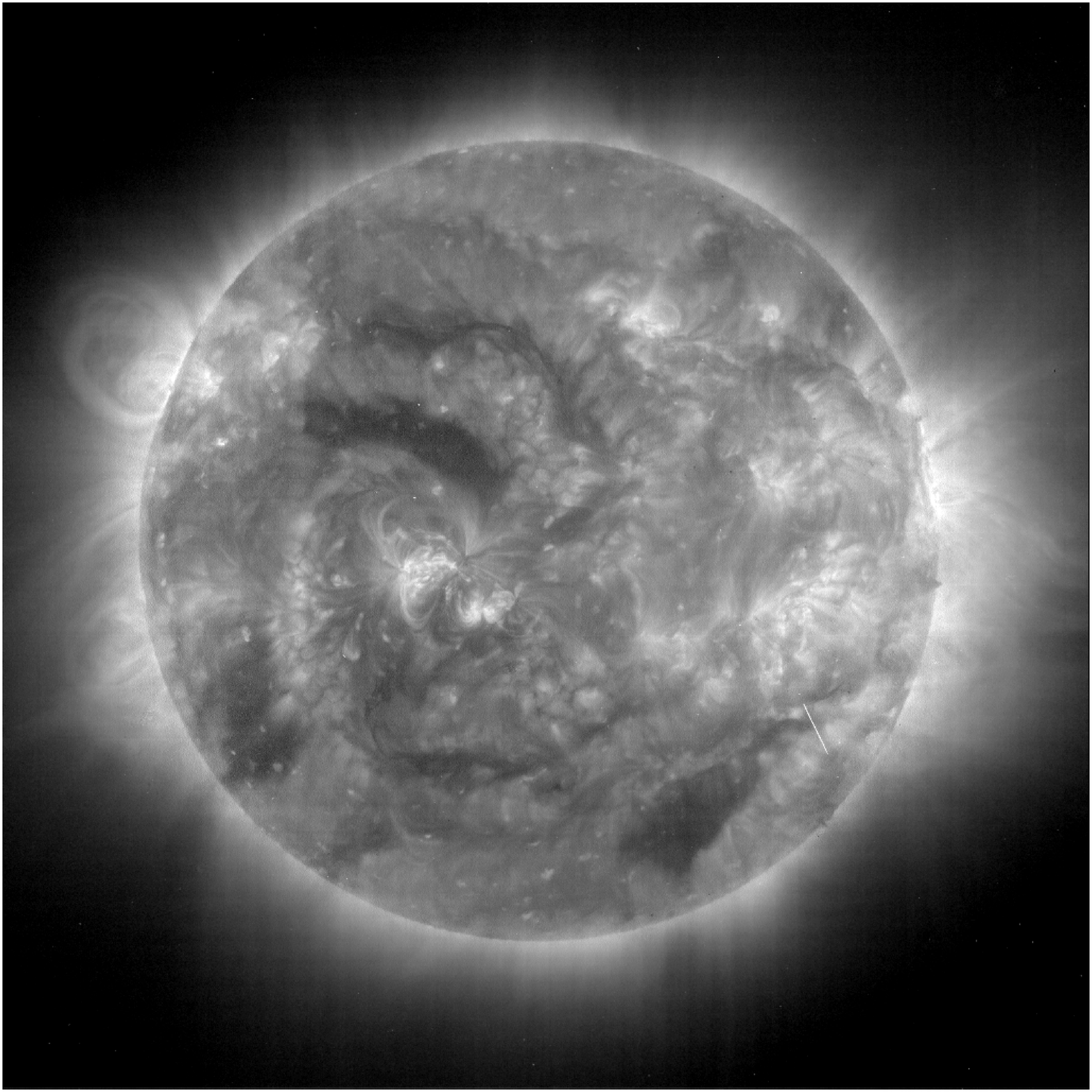}} &
    {\includegraphics[width=18pc,height=18pc]{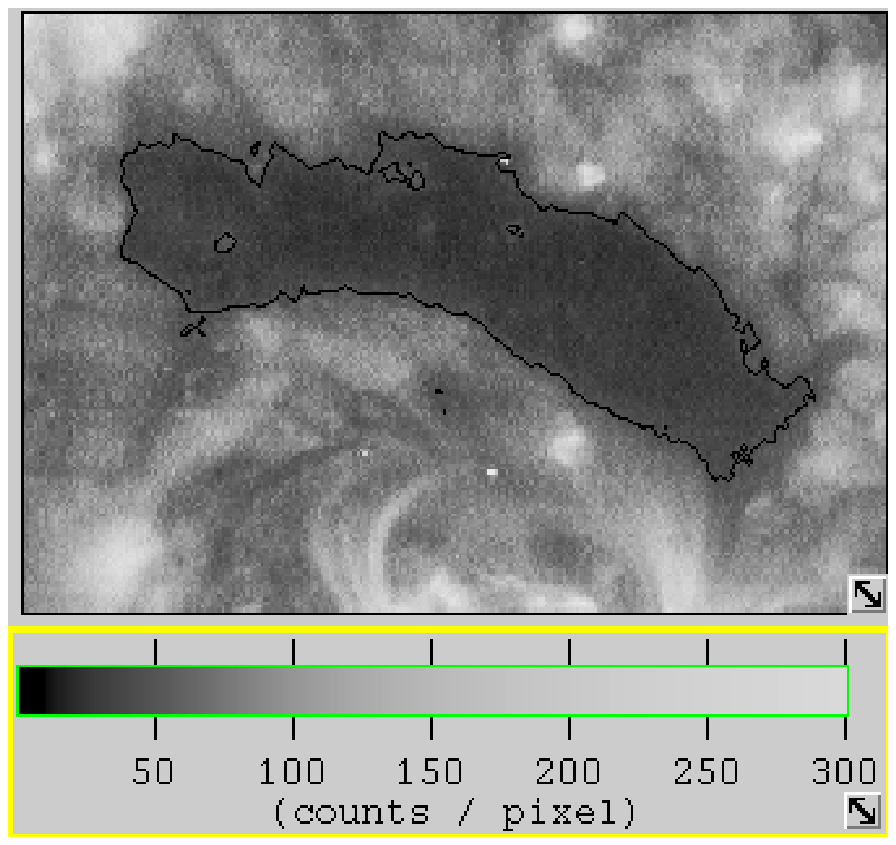}} \\
\end{tabular}

    \caption{Left side Fig 1 (a) shows full-disk SOHO/EIT 195 $\AA$ image of 01-01-2001, 00:24:11 UT with CH (in the north eastern hemisphere and close to center). This full disk image is kindly provided by Dr. Gurman.
Whereas, Fig 1(b) illustrates with a given threshold, contour map of the CH.}
\end{center}
 \end{figure}
 \begin{figure}
 \begin{center}
     Fig 2(a) \hskip 30ex Fig 2(b)
     \begin{tabular}{cc}
       {\includegraphics[width=16pc,height=16pc]{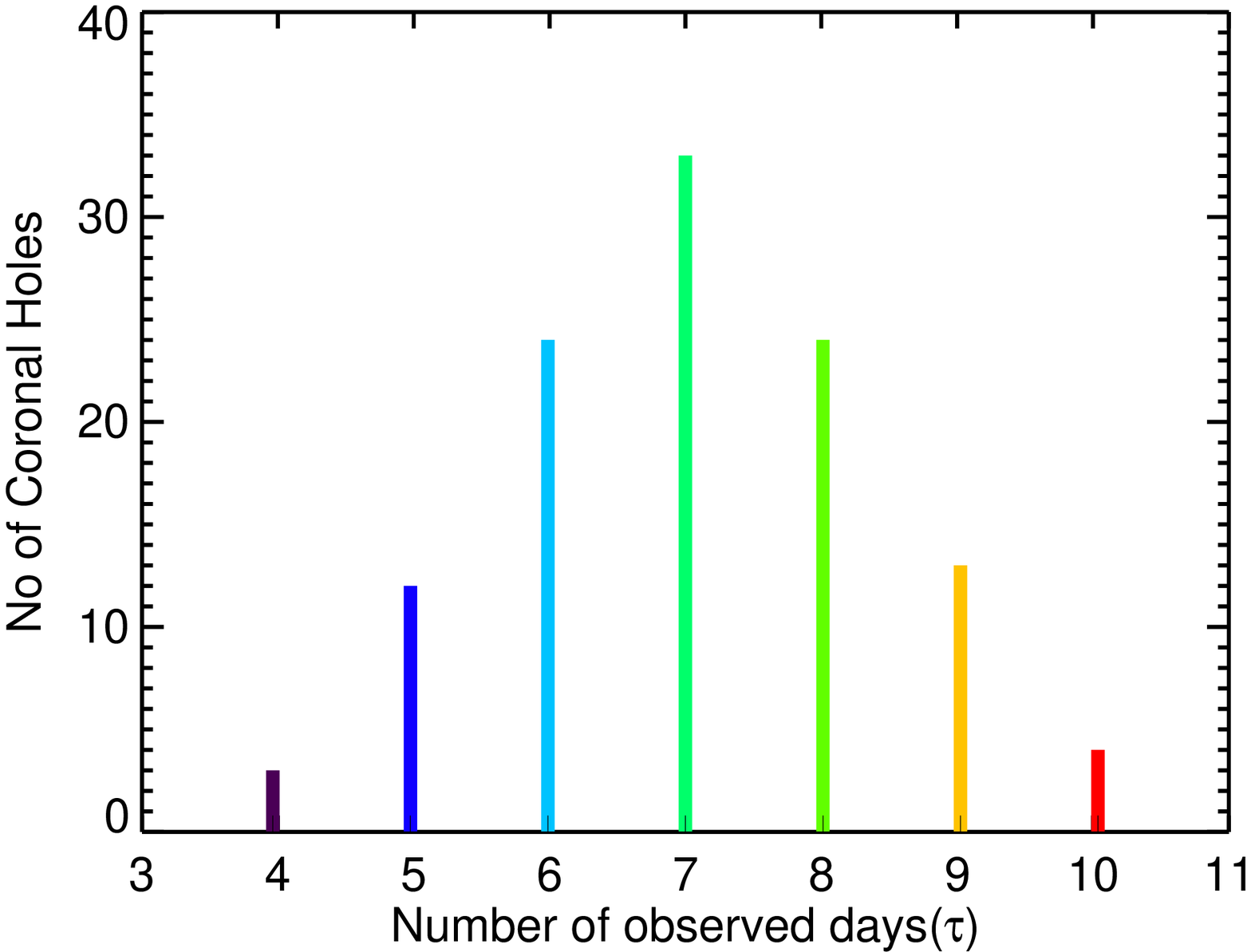}} 
  	{\includegraphics[width=16pc,height=16pc]{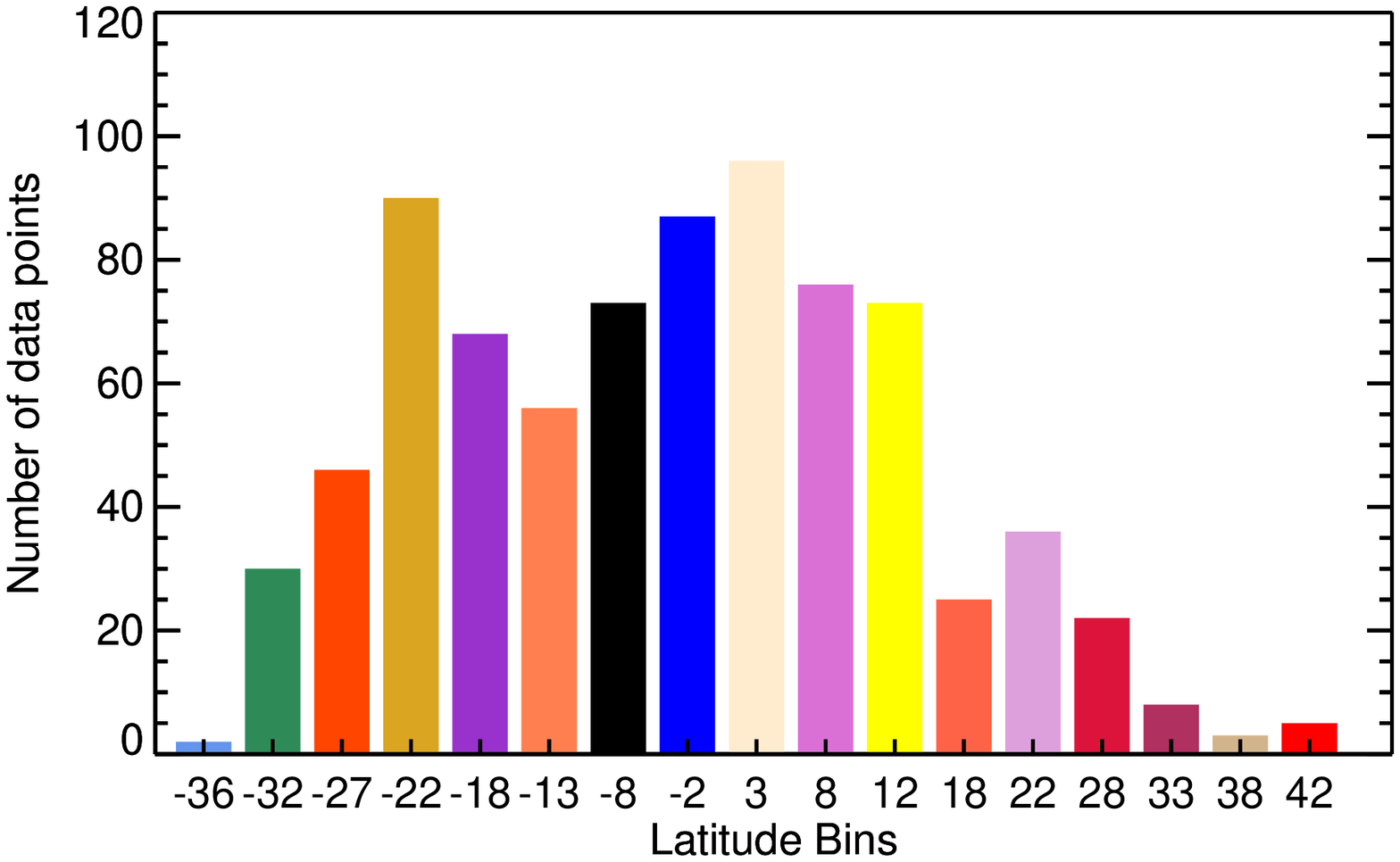}} \\
 \end{tabular}
 \caption{Figure 2(a) illustrates, number of coronal holes 
for different life spans on the solar disk.
Whereas, for different latitude bins in both the hemispheres, Fig 2(b) illustrates, number of coronal holes.}
\end{center}
 \end{figure}

\begin{figure}
 \begin{center}
     \begin{tabular}{cc}
       \includegraphics[width=40pc,height=36pc]{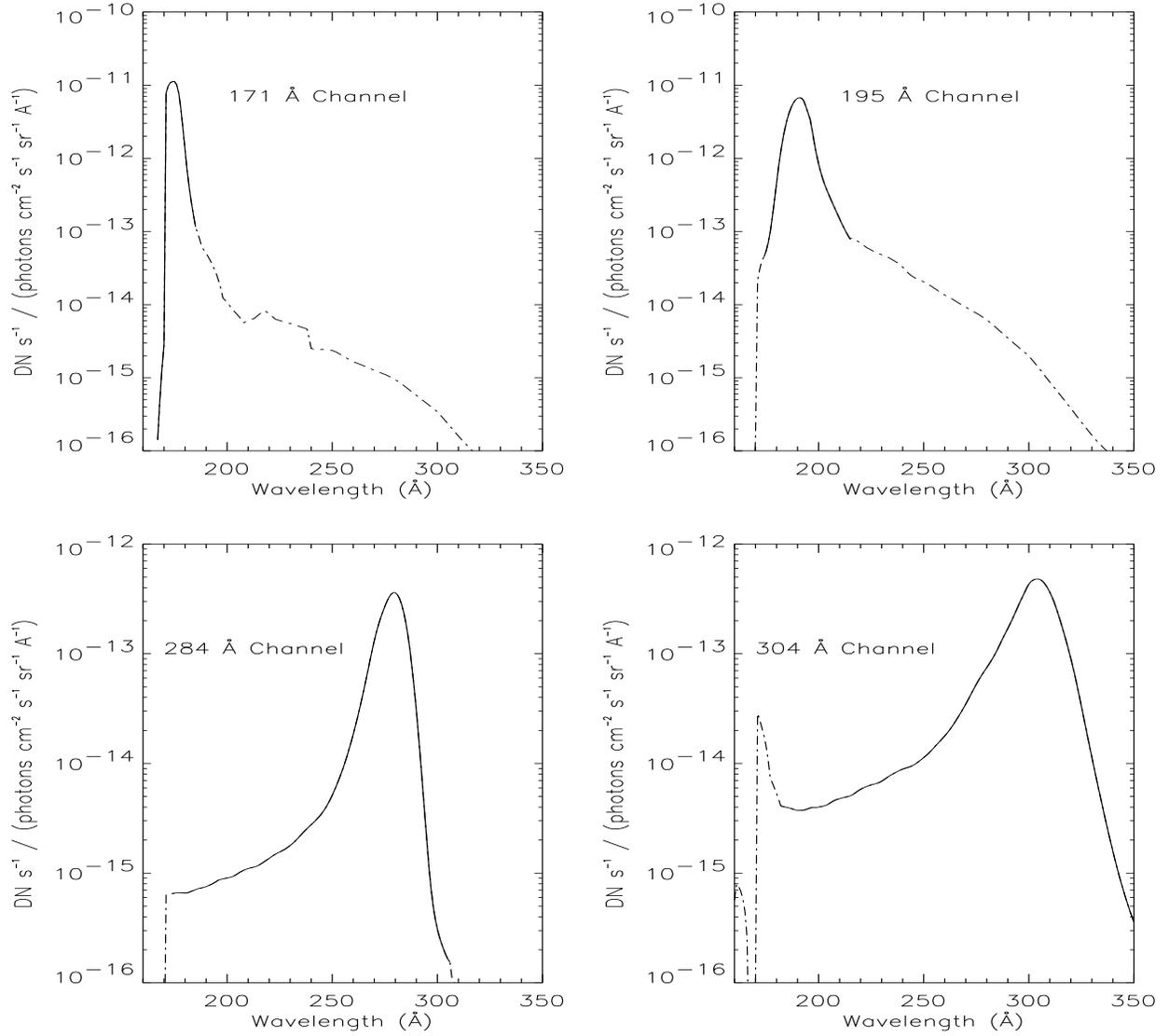} \\
 \end{tabular}
\caption{For different observed channels (171$\AA$, 195$\AA$, 284$\AA$ and 304$\AA$), this figure illustrates SOHO/EIT instrumental response curves.}
 \end{center}
 \end{figure}

\begin{figure}
    \begin{tabular}{cc}
      {\includegraphics[width=18pc,height=18pc]{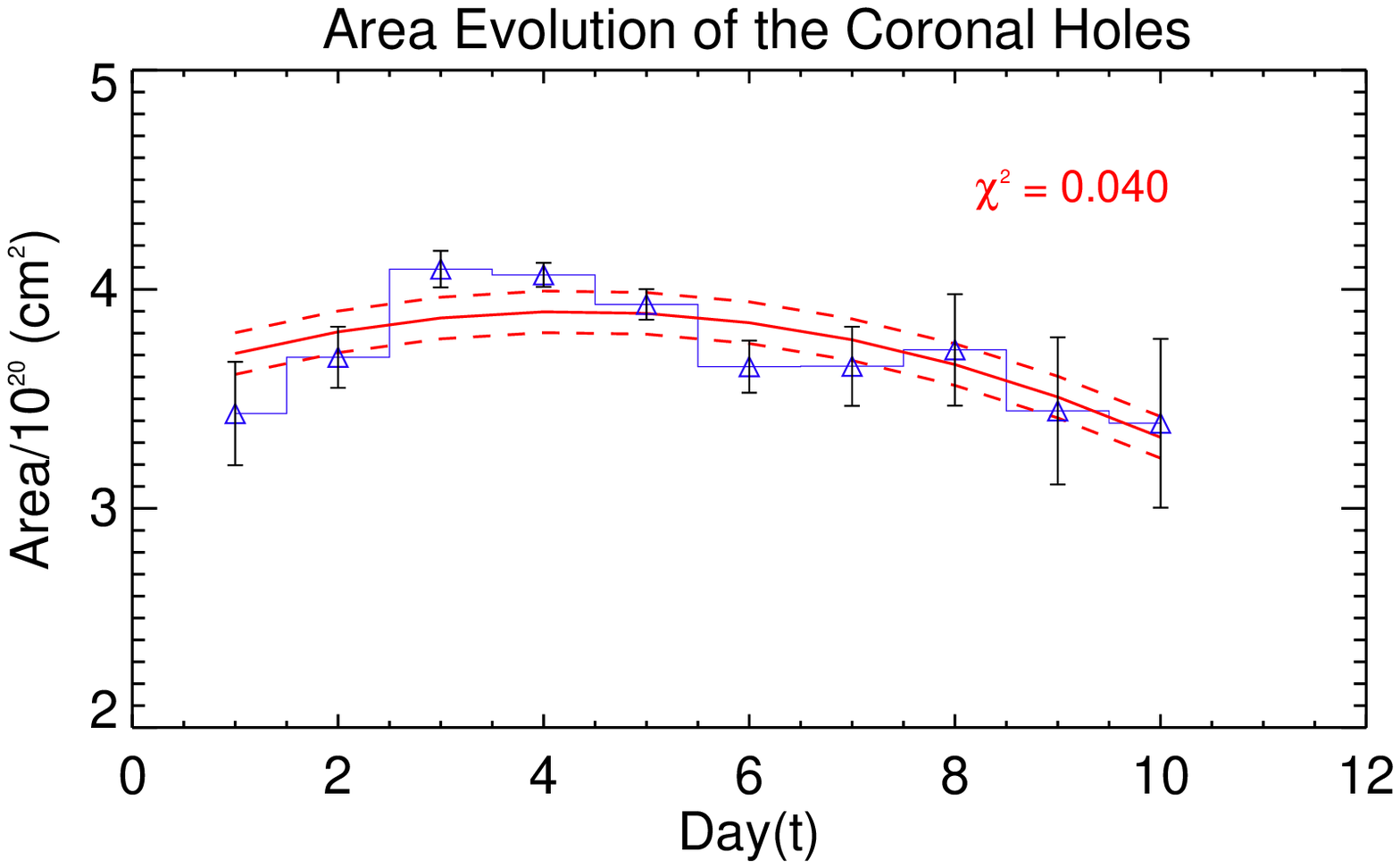}} &
      {\includegraphics[width=18pc,height=18pc]{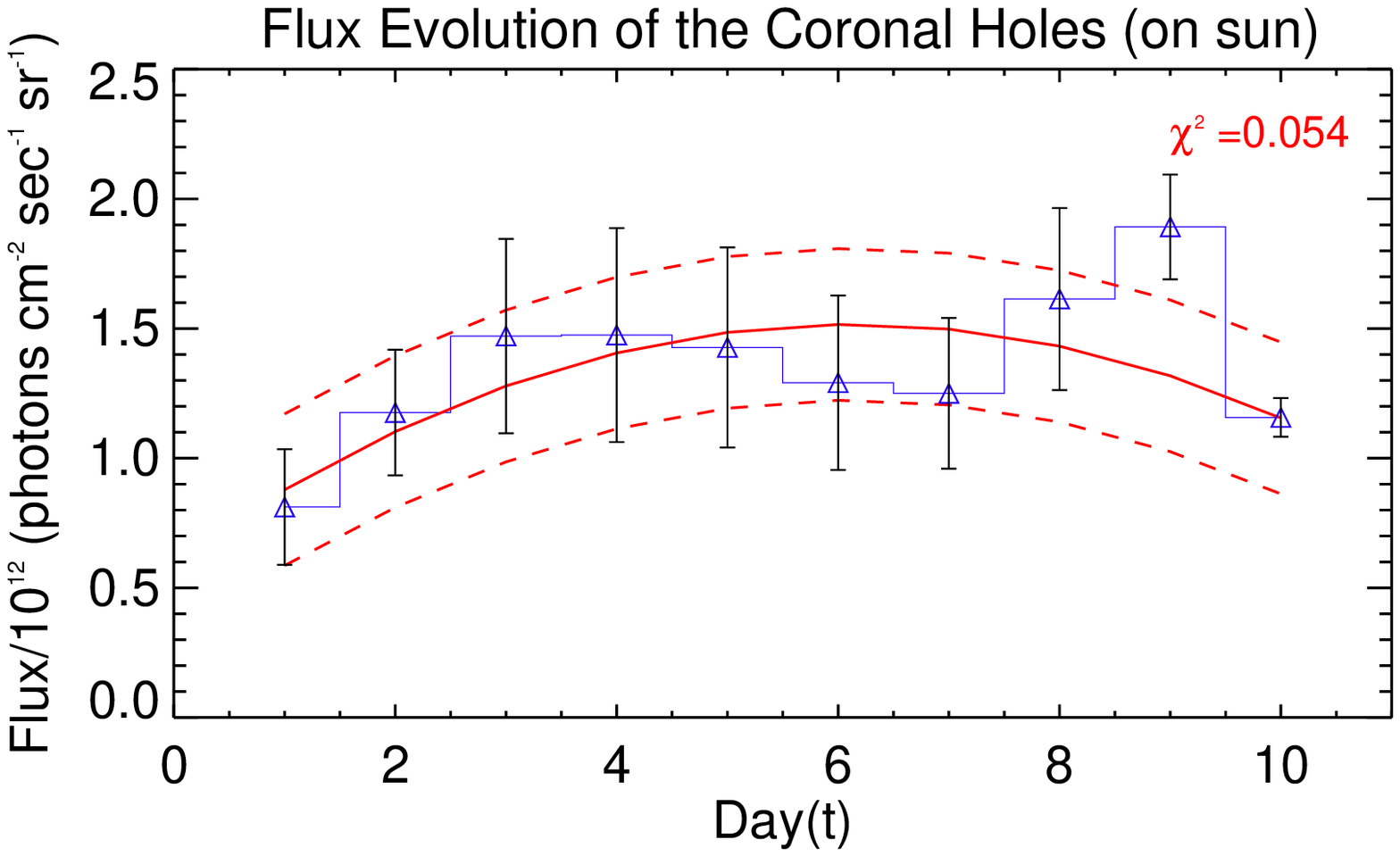}} \\
\end{tabular}

\begin{tabular}{cc}
      {\includegraphics[width=18pc,height=18pc]{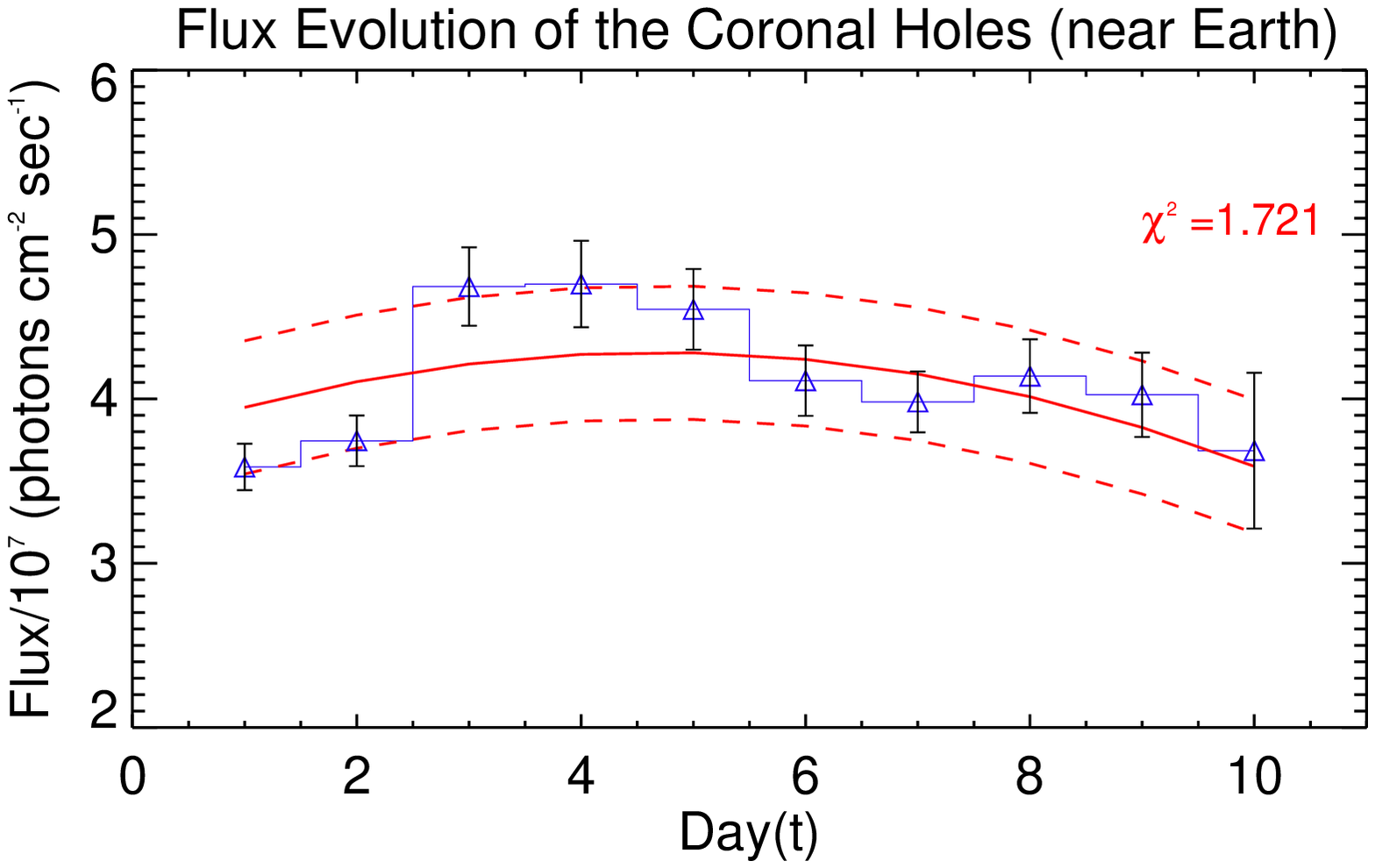}} &
     {\includegraphics[width=18pc,height=18pc]{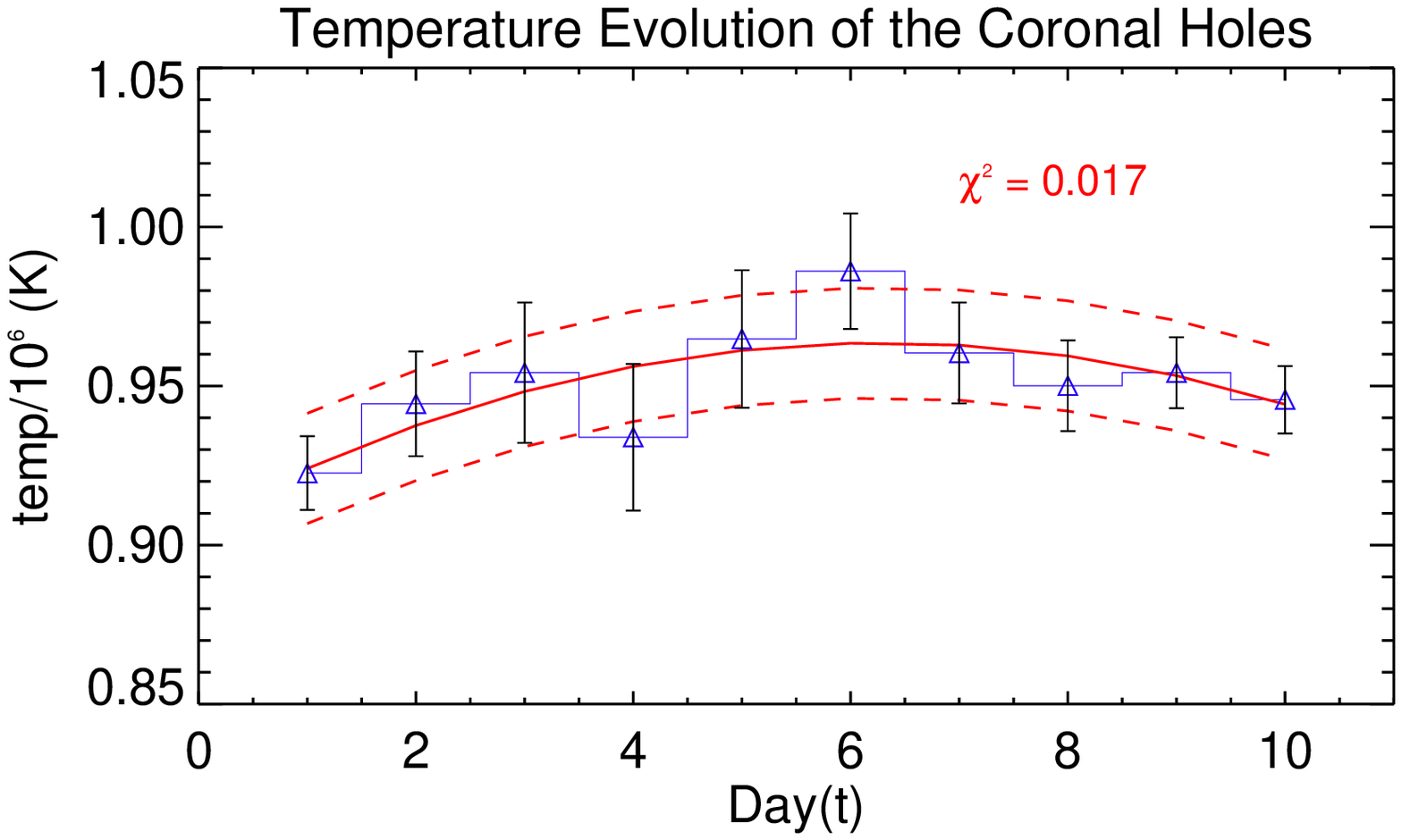}} \\
\end{tabular}
\caption{Irrespective of their latitudes, between +65$^\circ$ to -65$^\circ$
longitudes from the central meridian, variation of of different physical parameters of
coronal holes such as Area, Flux (on the sun and near Earth at 1 AU) and Temperature during
their evolution on the solar visible disk.
 In all the figures, blue bar plot  represents the observed values of Area, Flux, Temperature respectively and
the red continuous line represents a least-square fit
 $Y(t)=C_{0}+C_{1} t+ C_{2} t^{2}$ to the observed values.
 $Y(t)$ is the estimated different physical parameters,
$t$ is day of observation, and $C_{0}$, $C_{1}$ and $C_{2}$ are the constant coefficients
determined from the least square fit. Whereas the
red dashed lines represent the one standard deviation (which is computed
from all the data points) error bands. $\chi^{2}$ is a measure of goodness of fit.
}
\end{figure}

\begin{figure}
    \begin{tabular}{cc}
      {\includegraphics[width=18pc,height=18pc]{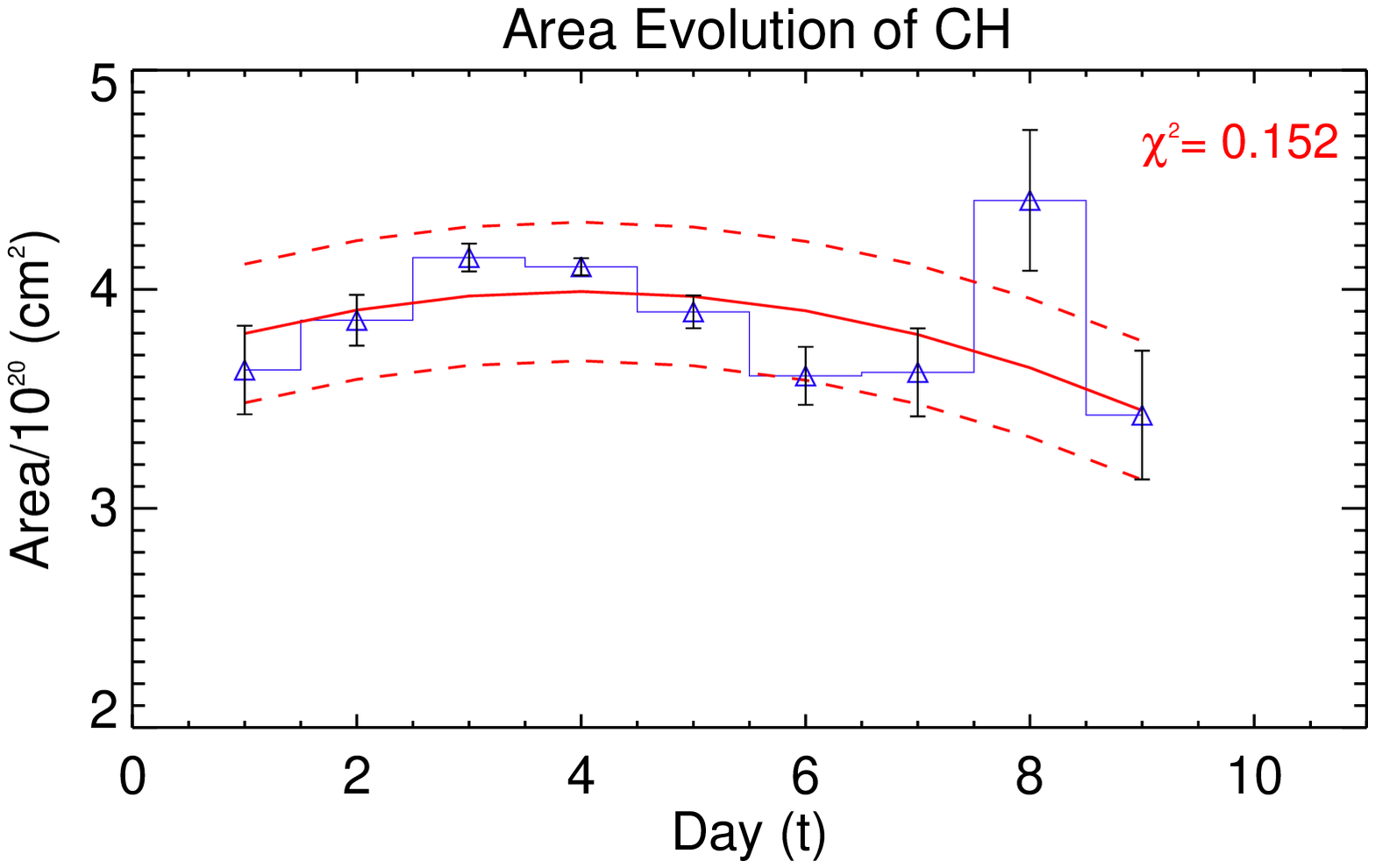}} &
      {\includegraphics[width=18pc,height=18pc]{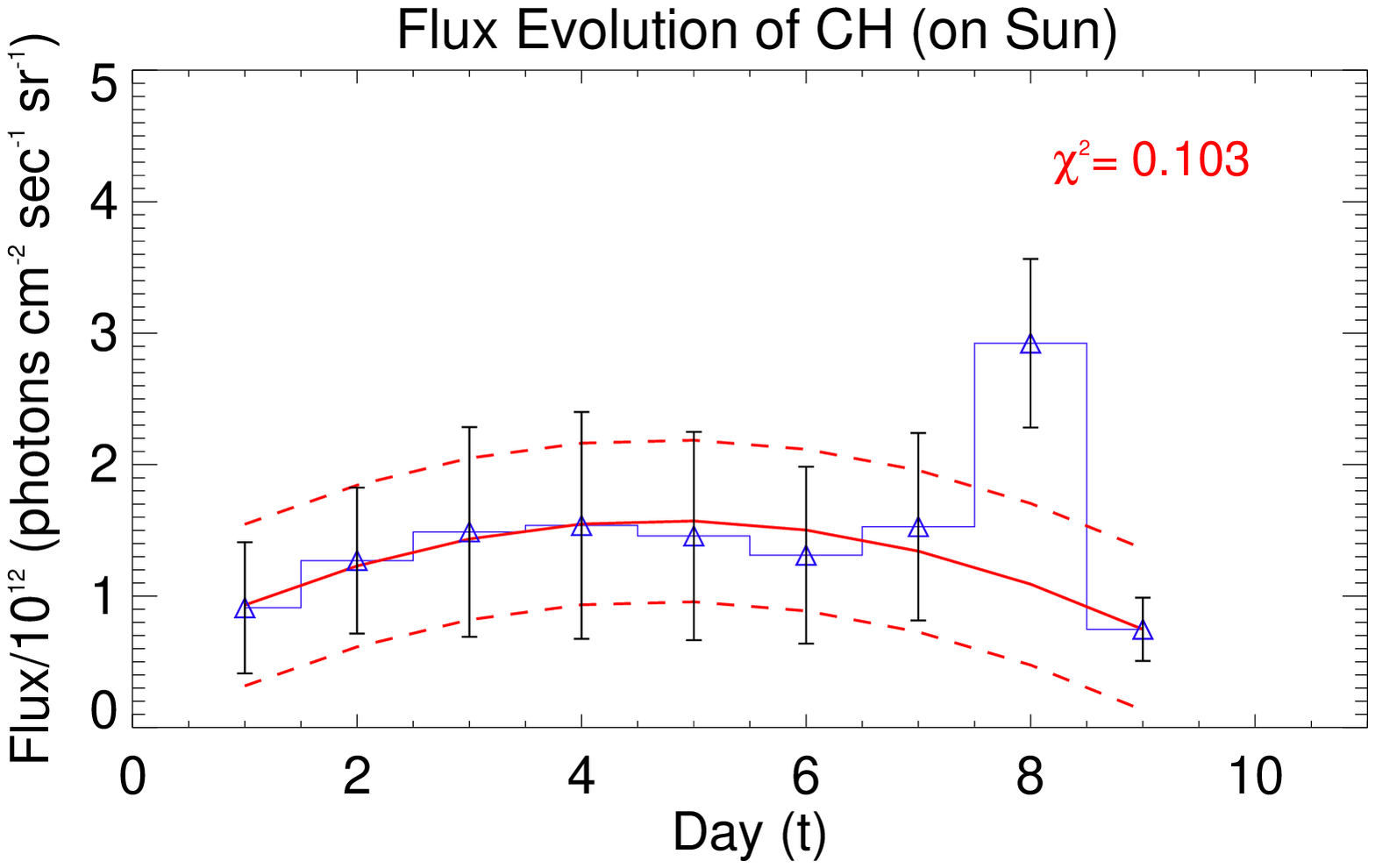}} \\
\end{tabular}

\begin{tabular}{cc}
      {\includegraphics[width=18pc,height=18pc]{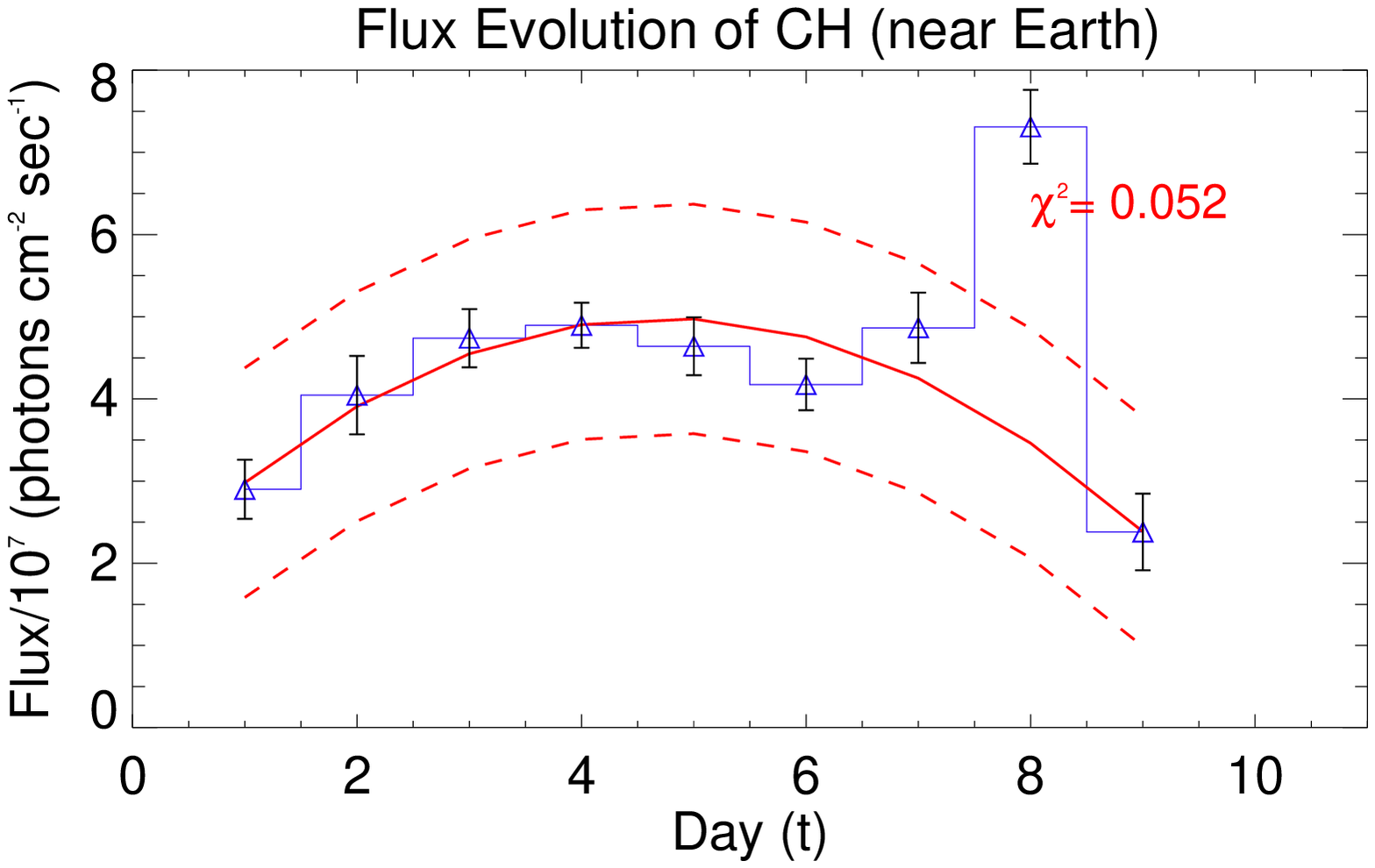}} &
      {\includegraphics[width=18pc,height=18pc]{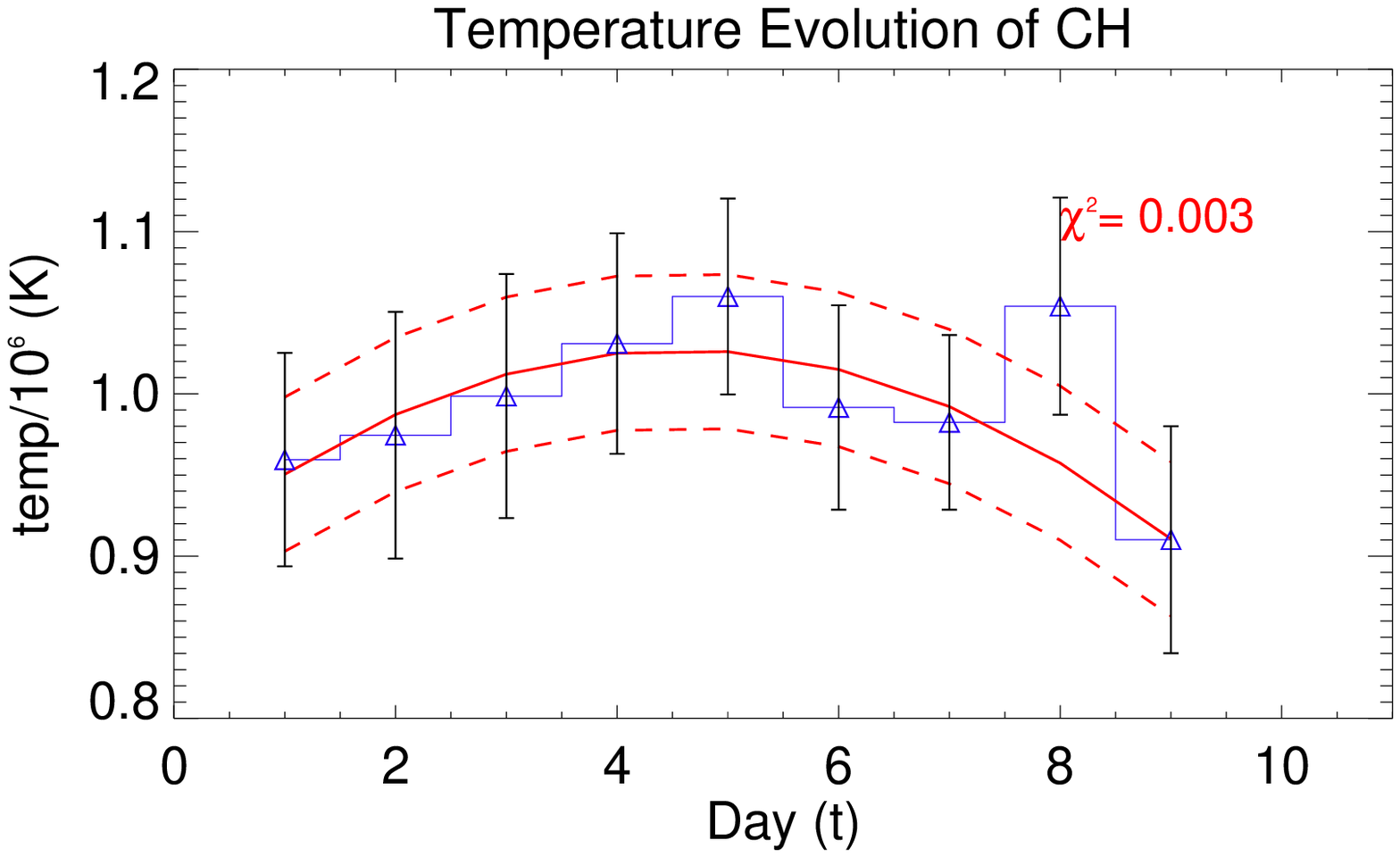}} \\
\end{tabular}
\caption{Irrespective of their latitudes, between +45$^\circ$ to -45$^\circ$
longitudes from the central meridian, daily variation of different physical parameters of
coronal holes such as Area, Flux (on the sun and near Earth at 1 AU) and Temperature during
their evolution on the solar visible disk.
 In all the figures, blue bar plot  represents the observed values of Area, Flux, Temperature respectively and
the red continuous line represents a least-square fit of the forms
 $Y(t)=C_{0}+C_{1} t+ C_{2} t^{2}$ to the observed values.
 $Y(t)$ is the estimated different physical parameters,
$t$ is day of observation, and $C_{0}$, $C_{1}$ and $C_{2}$ are the constant coefficients
determined from the least square fit. Whereas the
red dashed lines represent the one standard deviation (which is computed
from all the data points) error bands. $\chi^{2}$ is a measure of goodness of fit.
}
\end{figure}

\begin{table*}
\centering
\caption{Estimated CH parameters for +65$^\circ$ to -65$^\circ$}
 \begin{tabular}{@{}llllllllll@{}}
 \hline
Parameters & C$_0$ & $\delta$C$_0$ & C$_1$ & $\delta$C$_1$ & C$_2$ & $\delta$ C$_2$ & $\chi^{2}$ \\
\hline
A$\times10^{20}$ & 3.57 & 0.83 & 0.15 & 0.30 & -0.02 & 0.02 & 0.040\\
F$\times10^{11}$ & 6.06 & 0.30 & 2.97 & 0.70 & -0.20 & 0.06 & 0.054\\
F1$\times10^{7}$ & 3.70 & 0.90 & 0.23 & 0.10 & -0.03 & 0.01 & 1.721\\
T$\times10^{6}$ & 0.91 & 0.30 & 0.02 & 0.04 & -0.001 & 0.004 & 0.017\\
\hline
\end{tabular}
\end{table*}

\begin{table*}
\centering
\caption{Estimated CH parameters for +45$^\circ$ to -45$^\circ$}
 \begin{tabular}{@{}llllllllll@{}}
 \hline
Parameters & C$_0$ & $\delta$C$_0$ & C$_1$ & $\delta$C$_1$ & C$_2$ & $\delta$ C$_2$ & $\chi^{2}$ \\
\hline
A$\times10^{20}$ & 3.70 & 0.98 & 0.80 & 0.70 & -0.02 & 0.01 & 0.152\\
F$\times10^{11}$ & 5.40 & 1.20 & 4.30 & 0.90 & -0.50 & 0.09 & 0.103\\
F1$\times10^{7}$ & 1.80 & 1.80 & 1.30 & 0.90 & -0.14 & 0.09 & 0.052\\
T$\times10^{6}$ & 0.90 & 0.30 & 0.50 & 0.10 & -0.05 & 0.01 & 0.003\\
\hline
\end{tabular}
\end{table*}

\begin{table*}
\centering
\caption{Estimated CH parameters for +65$^\circ$ to -65$^\circ$}
 \begin{tabular}{@{}llllllllll@{}}
 \hline
Parameters & C$_0$ & $\delta$C$_0$ & C$_1$ & $\delta$C$_1$ & $\chi^{2}$ \\
\hline
A$\times10^{20}$ & 3.80 & 0.50 & -0.50 & 0.10 & 1.409\\
F$\times10^{12}$& 1.30 & 0.08 & 0.50 & 0.20 & 6.986\\
F1$\times10^{7}$ & 4.40 & 0.30 & 0.70 & 0.50 & 0.405\\
CT$\times10^{6}$ & 0.93 & 0.40 & 0.17 & 0.14 & 0.041\\
CP$\times10^{-2}$ & 0.97 & 0.20 & 0.17 & 0.15 & 0.041\\
Photospheric $|B|$ & 4.27 & 0.22 & 3.67 & 0.90 & 4.474\\
Corona $|B|$ & 0.075 & 0.020 & 0.09 & 0.04 & 0.610\\
MP$\times10^{-3}$& 0.18 & 0.04 & 0.77 & 0.15 & 2.478\\
TP$\times10^{-2}$ & 0.97 & 0.10 & 0.03 & 0.10 & 0.041\\
True temperature$\times10^{6}$ & 0.94 & 0.10 & 0.03 & 0.01 & 0.041\\
\hline
\end{tabular}
\end{table*}

\begin{table}
\centering
\caption{Estimated CH parameters for +45$^\circ$ to -45$^\circ$}
 \begin{tabular}{@{}llllllllll@{}}
 \hline
Parameters & C$_0$ & $\delta$C$_0$ & C$_1$ & $\delta$C$_1$ & $\chi^{2}$ \\
\hline
A$\times10^{20}$ & 3.80 & 0.50 & +0.50 & 0.10 & 1.409\\
F$\times10^{12}$ & 3.00 & 0.20 & 2.60 & 0.20 & 3.744\\
F1$\times10^{7}$ & 5.20 & 0.10 & 5.10 & 1.00 & 1.559\\
CT$\times10^{6}$ & 0.90 & 0.10 & 0.40 & 0.01 & 0.857\\
CP$\times10^{-2}$ & 0.93 & 0.10 & 0.50 & 0.10 & 0.857\\
Photospheric $|B|$ & 4.21 & 0.22 & 1.32 & 0.90 & 4.162\\
Corona $|B|$ & 0.08 & 0.01 & 0.03 & 0.01 & 4.162\\
MP$\times10^{-3}$& 2.40 & 0.10 & 3.40 & 0.20 & 0.032\\
TP$\times10^{-2}$& 0.01 & 0.001 & 0.0002 & 0.0001 & 2.481\\
True temperature$\times10^{6}$& 1.01 & 0.01 & 0.02 & 0.03 & 2.481\\
\hline
\end{tabular}
\end{table}

\section{RESULTS}
As in the previous study (Hiremath \& Hegde 2013), we follow the 
similar criteria in selection of coronal holes and adopt similar
method in binning the CH data for different latitudes. A typical coronal hole
in the north-eastern quadrant of the SOHO/EIT image
observed on 1st Jan 2001 is presented in Fig 1(a). Whereas separated coronal hole
with its boundary detected from the threshold criterion (see the details in 
section 2 of Hiremath
\& Hegde 2013) is presented in Fig 1(b). 
We define apparent life span $\tau$ (as actual life span is larger) as number
of observed days that have first and last appearance of CH on the same side of 
solar disk. In this way, for the observed coronal holes with different life spans
(minimum of 4 days to maximum of 10 days)
that appear at different latitudes and between +65$^\circ$ to -65$^\circ$ longitude
from the central meridian are illustrated in Fig 2(a). With the constraint that coronal holes
that occur between +40$^\circ$ (northern hemisphere) to -40$^\circ$ (southern 
hemisphere) latitude zones, 113 coronal holes with different
life spans ultimately yield totally 796 data points for the present analysis.  
As the minimum life span of coronal hole is 4 days, {\em transient coronal
holes} (that have life span $\le$ 2 days; Kahler and Hudson 2001) that might have different physical
properties are not included in this analysis. 
Irrespective of their life spans and after binning the coronal holes 
between 0-5$^\circ$,  5-10$^\circ$, etc., average latitudes of
CH are computed. Fig 2(b) illustrates the idea of how totally 796 observed
data points are distributed in different latitude bins.  

For the typical two coronal holes (first
CH is observed in the southern (latitude $\sim$ $30^{o}$) hemisphere and second CH
is observed in the northern (latitude $\sim$ $13^{o}$) hemisphere respectively),
following the methods presented in section 2, 
daily estimated different physical parameters are presented in Table 2. 
For the sake of comparison of temperature structure of CH computed
from our method, by employing filter ratio technique, we also computed 
temperature structure. For example, Hinode/XRT data of 12/02/2007 and 
15/08/2007 taken in Al-mesh/Al-poly filters are considered for temperature 
measurement. We followed the filter ratio method of 
Narukage et al. (2011) and temperature structure of these two
CH is estimated. We find that temperature structure of CH
estimated by our method and temperature structure of CH estimated by
method of Narukage et al. (2011) is of the same order. 
With this confidence in mind that both the methods yield the same temperature structures,
for all the 113 coronal holes considered for this study, we compute all 
the physical parameters (as defined in section 2) and are illustrated in 
Figs. 4-11. 

Our first objective is to understand daily evolution of coronal hole
during its passage over the observed solar disk. 
For example, irrespective of their life spans and latitudes, for longitudes
between +65$^\circ$ to -65$^\circ$,
that are combined together in both the latitudes, Fig 4. illustrates the
daily evolution of area (A), radiative flux (F) on the sun, 
radiative flux near the Earth (to be specific
at Lagrangian point $L_{1}$ where SOHO space craft is positioned) ($F_{1}$)
and temperature structure (T) of CH on the sun. Polynomial
 of degree 2 (quadratic) yields the best fit for the daily variation of CH. 
Different estimated parameters ($C_{0}$, $C_{1}$ and $C_{2}$) of the
polynomial fit with their respective uncertainties ($\delta C_{0}$, $\delta C_{1}$ and $\delta C_{2}$) and, $\chi^{2}$ (a measure of goodness of fit)
values are
presented in Table 3 (+65 to -65 deg longitudes) and in
 Table 4 (+45 to -45 deg longitudes) respectively. 

It is to be noted that , daily variations of radiative flux of CH 
may be useful for the Earth's 
ionospheric studies (Bauer 1973; Hinteregger 1976; Roble and Schmidtke 1979;  Richards, Fennelly, Torr 1994; Lilensten {\em et.al.} 2007; Dudok de Wita
and Watermanna 2010; Kretzschmar {\em et.al.} 2009). 
 According to Bauer (1973),
one of the principal ionizing radiation (others are solar x-ray photons,
galactic cosmic rays, solar cosmic ray protons, etc.)
 responsible for formation of planetary
ionospheres is solar EUV. Although all the EUV photons from sun's whole
disk can ionize the planetary atmospheres, EUV photons from coronal holes
have more momentum that probably disturb the ambient planetary ionospheres.
That means, probably transient behavior of EUV photons of the coronal 
holes leads
to transient or sudden disturbances in the planetary ionospheres.

 From observations of EUV images one can argue that 
much brighter surrounding coronal hole
must be responsible for ionizing the Earth's upper atmosphere
 rather than a dark coronal hole. However, there is a difference
in these two images. Compared to surrounding brighter region
of coronal holes, radiation emitted from the coronal hole is
mainly associated with fast solar wind, accompanied by increase
in temperature, is more effective in ionizing
the Earth's atmosphere and also leading to geomagnetic disturbances.

\subsection {Estimation Life Span of CH} 
As for solar physics point of view, it is interesting to estimate the actual
life span of coronal holes. If we consider daily variation of area of
CH on the solar disk, best fit yields the quadratic equation with
respect to time. As a first approximation, we can crudely estimate 
average life span of CH although individual coronal holes may have 
different life spans. After neglecting the non-linear term 
(as it is very small, see Figures 4 and 5 and, Table 3 and 4) and
assuming that daily variation of area curve is symmetric (with
respect to maximum area), for area
of CH to be zero when it decays completely, average life span 
is estimated to be $\sim$ 46 days.

\subsection {Estimation of Magnetic Diffusivity of Corona}
From this daily evolution area curve, we find the average area $\sim$ $10^{20}$
$cm^{2}$ from which average diameter (assuming that area $A$ is circular) $L$ 
of CH is estimated to be $\sim$ $6 \times 10^{9}$ cm.
Further making assumption that CH is a magnetic flux tube whose
area evolution is only dictated by magnetic diffusion, magnetic
diffusivity (=$L^{2}\over{\tau}$, where $\tau$ is life span) 
of the corona is estimated
to be $\sim \, 10^{13}$ $cm^{2}$/sec which is very close to estimates 
of previous studies
(Krista 2011; Krista {\em et.al.} 2011; Hiremath and Hegde 2013).
Similarly neglecting the second non-linear coefficient (which is not significant
as the error is of same order) in the empirical law of area
that is derived from the least square fit (Fig. 4), one can estimate the
area derivative $dA\over {dt}$ and by employing the derived
area derivative law ($dA\over {dt}$=$\eta {d^2 A \over{dr^{2}}}$)
from the previous study (Hiremath and Hegde 2013;
see the section 4), magnetic diffusivity $\eta$ of the corona is estimated
to be of similar order ($10^{13}$ $cm^{2}$/sec), as estimated above.
Interestingly and it is to be noted that magnetic diffusivity
estimated by our both the methods and magnetic diffusivity estimated
by the previous studies (Krista 2011; Krista {\em et.al.} 2011) 
are of the same order.

\subsection{Thermal and Magnetic Field Structures of the Coronal Holes} 
Knowing thermal structure is very important
in understanding the fast solar wind emanating from the
coronal hole (Hegde et al. 2015 and references there in). 
It is also important to examine whether
thermal structure of coronal is independent
or dependent on the latitude which in turn may give clue as
to why high latitude coronal holes have high solar wind
velocity compared to the low latitude or equatorial
coronal holes (McComas, Elliott and Steiger 2002) 
during the minimum period. It is also interesting to know whether
from thermal pressure of coronal hole 
(if CH is a magnetic flux tube, the CH pressure
is combination of plasma and magnetic pressure) 
one can separate
the magnitude of average magnetic field structure of
coronal holes at the corona. These important observed physical parameters
such as thermal and magnetic structures are essential to
probe the depth (Hiremath and Hegde 2013) and structure
of the coronal holes. In the following study, first by
estimating the radiative flux, with the assumption of
thermodynamic equilibrium, we estimate the temperature
structure. This temperature structure we call 
as {\em total temperature}. This assumption of thermodynamic
equilibrium is almost akin to the same assumption employed
in differential emission measure (for example Hahn, Landi and Savin 2011 )
in estimating temperature structure of coronal holes.
Further by using the observed  density of CH, we compute the 
{\em total pressure}. The obvious reason for calling {\em total pressure} 
is that CH is a magnetic flux tube whose total pressure
is the combination of thermal and magnetic pressures.
 This concept basically is invoked from the Parker's 
(1955) idea wherein if magnetic flux tube is in
hydrostatic equilibrium, a combination of gas and
magnetic pressure in the flux tube is balanced
by the external gas thermal pressure.
In the following section, by using information of estimated 
total temperature and pressure, we not only derive the actual temperature
structure but also estimate the average magnetic field
structure of the coronal hole.
   
\subsubsection{Latitudinal Variation of Thermal Structure of Coronal Hole}

Many previous studies (Zhang et al. 2007; Landi 2008 etc.) concentrated
on single coronal hole and estimated average temperature
structure. As for our knowledge, this is a first study which
uses many coronal holes for estimating the thermal structure.
Following the previous study (Hiremath and Hegde 2013) in
selecting the data, first we estimate the different
physical parameters of CH that occur between +65$^\circ$ to -65$^\circ$
longitudes from the central meridian. 
Following the method described in section 2
and irrespective of their areas and number of observed days
on the solar disk, for different latitudes, variation
of different physical parameters such as area, radiative
flux (on the sun and near Earth) and total temperature
structure are presented in Fig 6. For example, one can notice
from the least-square fit (of the form 
$Y(\theta)=C_{0}+C_{1} sin^{2}\theta$, where $\theta$ is latitude,
$C_{0}$ and $C_{1}$ constants and, Y represents 
 different physical parameters) of area-latitude curve that, on average,
equatorial coronal holes have large area compared
to high latitude coronal holes. Whereas other illustrations
of radiative flux and temperature structures follow inverse
latitudinal variations, rather than 
following similar area-latitude relationship.
Inconsistency of both these results probably could
be due to contribution of coronal holes that are near
+65$^\circ$ to -65$^\circ$ longitudes, close to the limb. Hence,
in order to completely minimize such projectional effects,
further we restricted the data of CH that occur between
+45$^\circ$ to -45$^\circ$ longitudes from the central meridian
and the same results that are presented in Fig
6 are also presented in Fig 9. 
 It is to be noted that although
there is more variation, especially in case of Figures 9(c) and 9(d),
over all trend of least square fits for all the Figures is same
(of the form $c_{1}+c_{2}sin^{2} \theta$).
Whereas, in case of illustrations 6, except Figures 6(b)-6(d),
law of least square fit for Figure 6(a) is different
(of the form $c_{1}-c_{2}sin^{2} \theta$).
However, one can notice from Fig 9
that inconsistency in latitudinal variation of 
area and flux curves is removed and
all the illustrations have same latitudinal variations. 

Latitudinal variation of different physical parameters of CH
are subjected to least square fit (of the the form 
$Y(\theta)=C_{0} + C_{1} sin^{2} \theta $) and estimated coefficients
 $C_{0}$ and $C_{1}$
with their respective uncertainties and $\chi^{2}$ values 
are presented in Tables 5 and 6 respectively.
In both the tables, with the units in cgs, first column
consists of area (A), radiative flux (F) on the sun, flux (F1) near earth,
total temperature (CT), total pressure (CP), strength
of coronal hole magnetic field structure at the photosphere
 ($|B|$) and at the corona ($|B_{c}|$) and, magnetic pressure (MP) of
the coronal hole at corona, actual pressure (TP) and
temperature of the coronal hole at the corona. Whereas second
to fifth columns represent the coefficients and their
uncertainties as estimated by the least square fit.
The last column in both the tables represents the value
of $\chi^{2}$ (a measure of goodness of fit).

 Although variation
of thermal structure (for example observed inferences:
David et.al. 1998; Landi 2008; Landi and Cranmer 2009;
Hahn, Landi and Savin 2011 and, theoretical
inferences: Osherovish et.al. 1985) of the coronal hole at different
heights in the corona is available, studies on
latitudinal variation of thermal structure of
coronal holes are not available. In this respect,
we believe that, our
results on latitudinal variation of thermal structure
of coronal holes will be very useful to the solar community.

\subsubsection{Estimation of Strength of Magnetic Field Structure 
of the Coronal Holes}

From the latitudinal variation of total temperature and the observed
density  structure (Doschek et al. 1997 etc.) of coronal holes, with respect
to latitude, total pressure $P$ ($=2n_{e}kT$, where $n_{e}$ 
is number of electron
density, $k$ is Boltzmann constant and $T$ is estimated
total temperature; it is assumed that coronal hole plasma
has same number of electrons and protons density hence the
number $2$ is multiplied) is estimated and is presented in Fig 7(a) (for
the +65$^\circ$ to -65$^\circ$ longitudes). In order to minimize the 
projectional effects, same parameter
(for the +45$^\circ$ to -45$^\circ$ longitudes 
from the central meridian) is presented in Fig 10(a).
One can notice from both the figures that latitudinal
variation of total pressure of the coronal hole depends upon
the latitude such that equatorial coronal holes have low
pressure compared to high latitude coronal holes. In case one
accepts that coronal hole is a magnetic flux tube, then total
pressure in the coronal hole is sum of plasma and magnetic pressures. 
Offcourse, best analogy (except strong magnetic fields in sunspots) between
magnetic flux tubes (sunspots) and coronal hole can be obtained from
the previous (Fla {\em et.al.} 1984; Davila 1985; Cally 1986, 1987; 
Osherovich et.al., 1985; Ofman 2005 and references
there in; Obridko and Solovev, 2011)
MHD models.

Hence, if we accept that plasma pressure of coronal hole
is independent of latitudes, then one possible interpretation 
for latitudinal variation of total pressure of the coronal holes could be
due to contribution from the magnetic pressure. That means
if one knows the latitudinal variation of magnetic field
structure of coronal holes, one can compute the magnetic
pressure and can be subtracted from the estimated total
pressure. Unfortunately, to the knowledge of the authors, 
there are no such studies that give the information of latitudinal variation
of magnetic field structure of coronal holes at the corona. For this
purpose, we adopt the following method in computing the
magnetic field structure and hence magnetic pressure of
the coronal hole at the corona.

First we make the reasonable assumption that coronal hole
is a magnetic flux tube that probably anchored below the
photosphere (Gilman 1977;
Golub {\em et.al} 1981; Jones 2005; 
Hiremath and Hegde 2013). As the magnetic flux tube is
embedded in the solar atmosphere, increase with height
from photosphere to corona results in decrease of 
surrounding ambient plasma pressure and hence tube must expand.
In the following, this statement can further be corroborated from the
previous study (Hegde, Hiremath and Doddamani 2014).
 With the SDO data, for the coronal
hole observed in three wavelengths 174 $\AA$, 193 $\AA$ and
211 $\AA$, average area of
coronal hole are :  $0.5 \times 10^{20}$ $cm^{2}$ for 174 $\AA$, $0.98 \times 10^{20}$ 
$cm^{2}$ for 193 $\AA$ and, $1.06 \times 10^{20}$ $cm^{2}$ for 211 $\AA$ respectively. 
Simultaneously DN counts (radiative intensity) also reduce from 174 $\AA$, 193 $\AA$
and 211 $\AA$ respectively. Successive increase of line formation
(Yang {\em et.al.} 2009) for 174 $\AA$, 193 $\AA$ and 211 $\AA$ at different heights are:
1.01$R_{\odot}$, 1.05$R_{\odot}$ and 1.3$R_{\odot}$. One can notice
that within the 30\% of solar radius from the photosphere,
coronal hole's area increases twice the area of coronal
hole at the photosphere. Considering these facts it is reasonable
to consider the coronal hole is expanding from
photosphere to corona where 195 $\AA$ line originates.

Further if one accepts that coronal hole is a Parker's
 flux tube (Parker 1955), then magnitude of magnetic
field structure $B$ inside the coronal hole is directly
proportional to ${P_{e}}^{1/2}$ (where ${P_{e}}$ is external ambient
pressure of the plasma). Hence, with this simple relationship,
one can estimate strength of magnetic field structure $B_{c}$ of
the coronal hole at the corona if one knows the strength of 
magnetic field structure and ambient plasma pressures at different heights.
To be specific, the resulting derivation will be (that is
obtained from the above simple relationship) 
B$_{c}$=B$_{pho}$ ($P_{ce} \over P_{pho}$)$^{1/2}$
(where $B_{c}$ and $P_{ce}$ are strength of magnetic field structure
of the coronal hole and ambient plasma pressure at the corona and, 
P$_{pho}$ and B$_{pho}$ are strength of magnetic field structure
of the coronal hole and ambient plasma pressure at the photosphere).  
As for strength of magnetic field structure $B_{pho}$ of coronal hole
at the photosphere, for the same time of observation
and latitude, we consider the inferred values
at the photosphere (using Solar Monitor website). 
 This inferred field from the photospheric magnetograms
is the line of sight component.
Then with the above formula and, 
by using the ambient external pressure at the photosphere 
and the corona (Aschwanden 2004), magnetic
field structure $B_{c}$ (and hence magnetic 
pressure {$B_{c}^{2} \over {4\pi}$}) 
of the coronal hole at the
corona (around 1.1 $R_{\odot}$, where the 195 $\AA$ line is
originated) is estimated. 

  After binning in different latitude
zones, latitudinal variation of average strength of 
magnetic field structure of coronal
hole at the photosphere are presented
in Fig 7b (for +65$^\circ$ to -65$^\circ$ longitudes) and
in Fig 10b (for +45$^\circ$ to -45$^\circ$ longitudes). It is interesting
to note that, as we reasoned above, indeed magnitude
of magnetic field structure (and magnetic pressure) of 
the coronal hole at the photosphere increases
from equator to higher latitudes  although curve of
Fig 10(b) appears to be flatter. Estimated strength of magnetic
field structure of the coronal hole and its magnetic pressure
at the corona are presented in Figures 7(c) and 7(d) 
(for +65$^\circ$ to -65$^\circ$ longitudes) and in Figures 10(c) and 
10 (d) (for +45$^\circ$ to -45$^\circ$ longitudes). One can notice
that, on average, magnitude of magnetic field structure of
coronal hole at the corona is estimated to be $\sim$ 0.08($\pm$0.02) Gauss.  

 Offcourse, one can argue from the spatially resolved individual coronal
holes that this estimated field strength depends
on whether the coronal holes are in new active regions where the fields are
strong or in old expanded unipolar regions where the fields are weak.
However, our estimated strength of magnetic field structure
is derived from the least square fit (first coefficient of the linear
 fit) with many coronal holes  rather than
obtaining the spatial information.
 
 There are also other
possibilities that this estimated strength of magnetic
field is possibly different than the actual strength
of magnetic field structure for the following two reasons.
Firstly observed magnetic field structure of the coronal hole
estimated from the photospheric magnetogram is longitudinal and hence
inferred magnetic field of the coronal hole at corona
is also a longitudinal component. Whereas for the radial component
of magnetic field (such a field structure is invoked in modeling of
coronal hole), strength of longitudinal component of magnetic 
field of coronal hole appears to be under
estimated as radial field is $B_{l}/cos(\lambda-\phi)$
(where $B_{l}$ is longitudinal component of magnetic
field, $\lambda$ is latitude (that varies 0 to 90 deg
from equator to pole) and $\phi$ is inclination angle of rotational
axis of the sun). Secondly and according to
Parker's (1955) Flux tube model, estimated strength of magnetic field depends
upon square root of ambient pressure which ultimately is model
dependent. Hence, it can not be ruled out that some amount
of uncertainty (0.01, $\sim$ $13\%$) persists that is reflected
in the estimated strength of longitudinal component of 
magnetic field from the least square fit.
However, it is important
to be noted that we haven't come across any study that estimates
strength of magnetic field of the coronal hole observed in
195 $\AA$.

By knowing latitudinal variation of strength of magnetic field
structure $B$ of the coronal hole, the estimated magnetic 
pressure is subtracted from the total
pressure of the coronal hole and actual thermal pressure of the
coronal hole is computed. By knowing electron density and
thermal pressure (Figures 8(a) and 11(a)), actual temperature 
structure of corona hole
is computed and latitudinal variation of the same is
presented in Fig 8(b) (for +65$^\circ$ to -65$^\circ$ longitudes)
and in Fig 11(b) (for +45$^\circ$ to -45$^\circ$ longitudes) respectively.
From these figures, we find that variation of temperature structure
of coronal holes is independent of solar latitudes.

\begin{figure}
\begin{center}
 \hskip 8ex 6(a) \hskip 40ex 6(b)
    \begin{tabular}{cc}
      {\includegraphics[width=18pc,height=18pc]{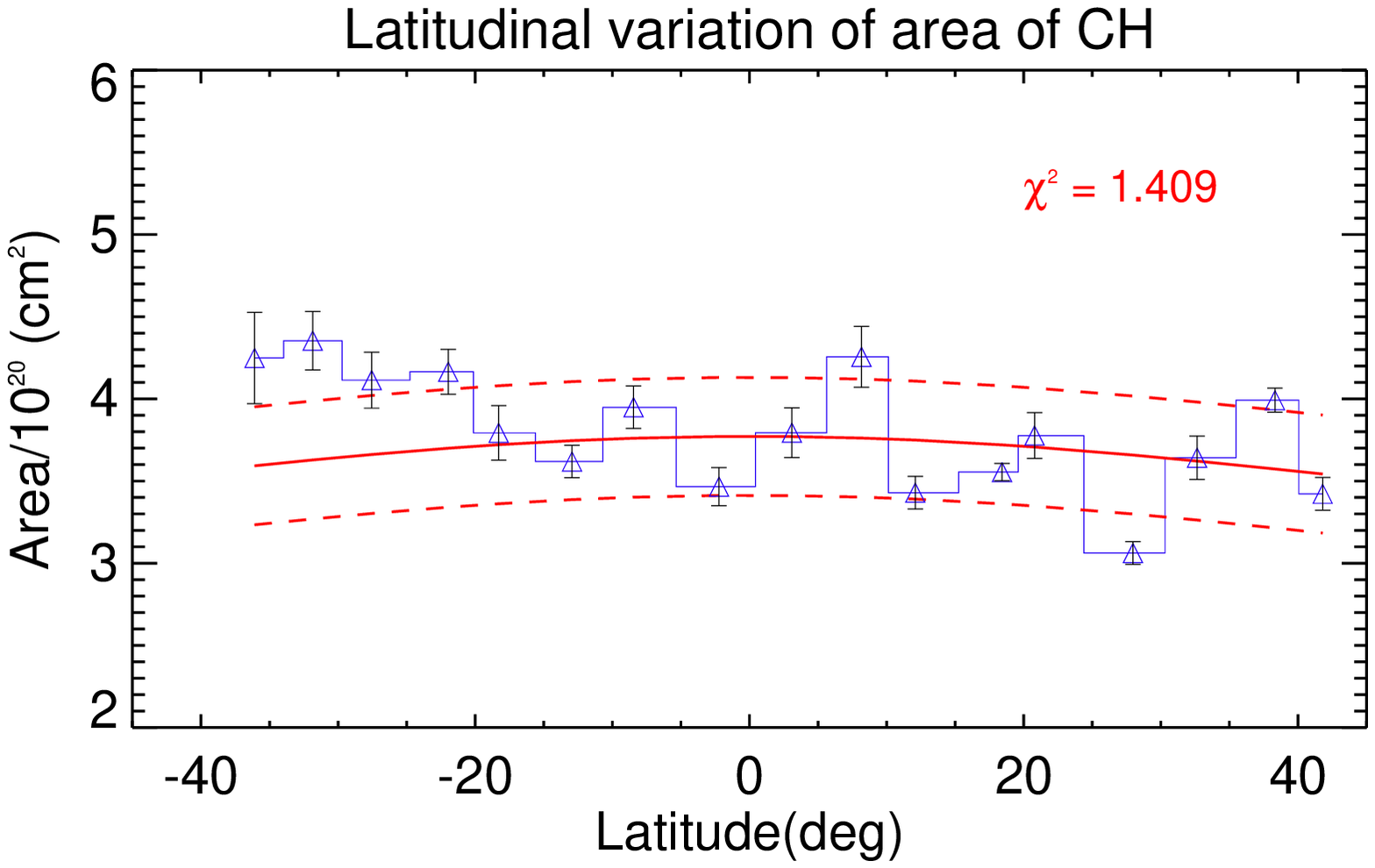}} &
      {\includegraphics[width=18pc,height=18pc]{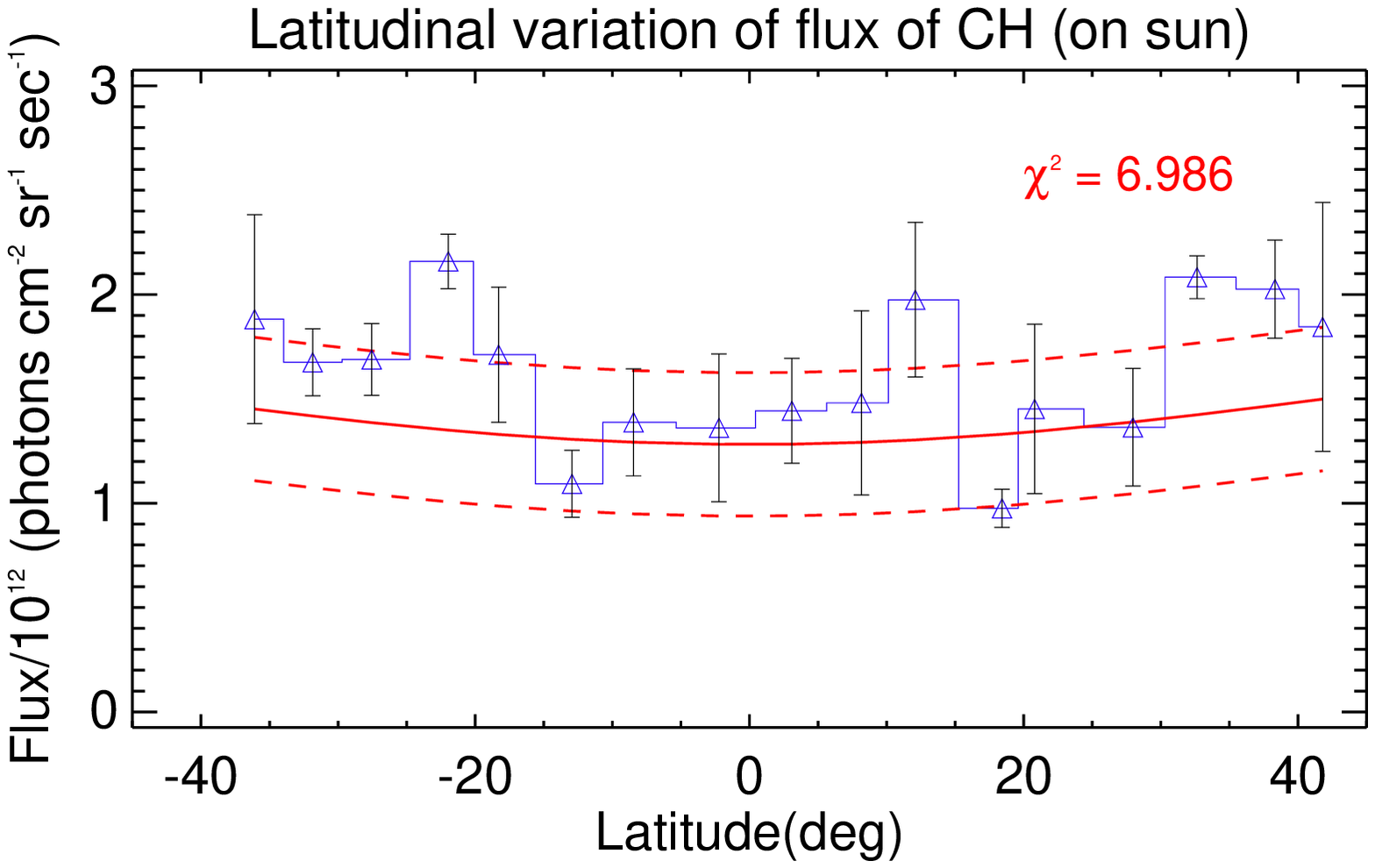}} \\
\end{tabular} 

\begin{tabular}{cc}
      {\includegraphics[width=18pc,height=18pc]{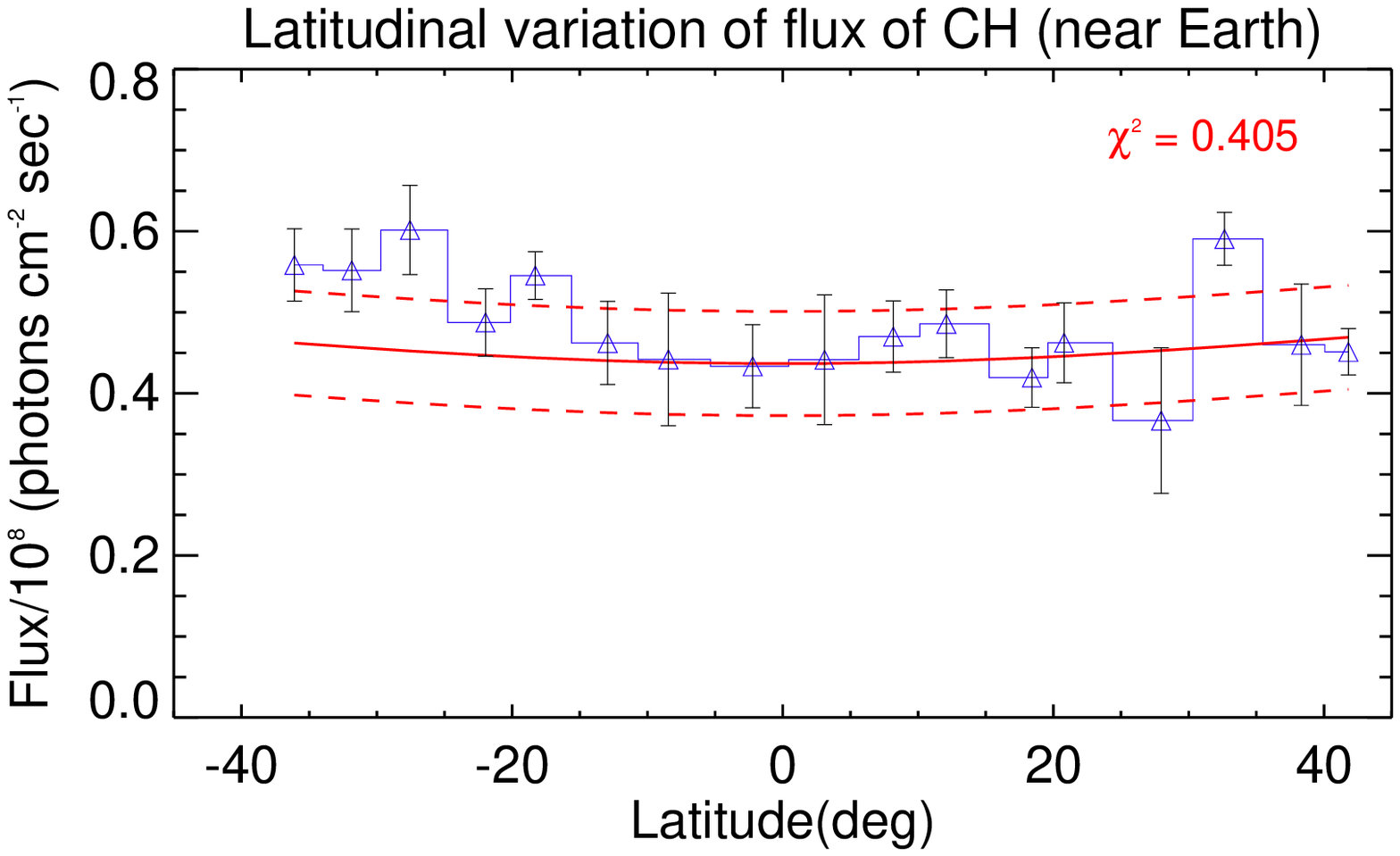}} &
{\includegraphics[width=18pc,height=18pc]{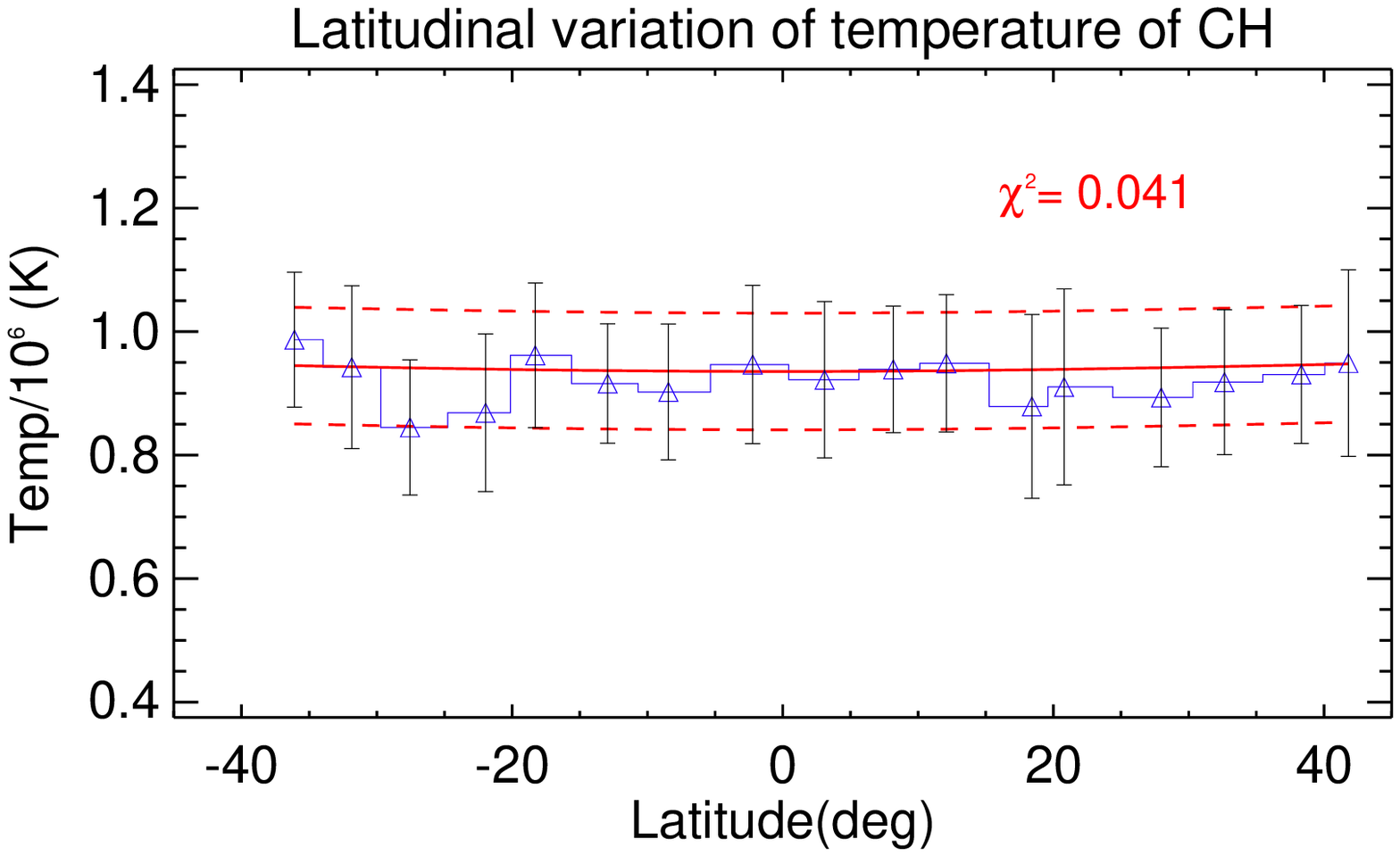}} \\
\end{tabular} 
\hskip 8ex 6(c) \hskip 40ex 6(d)
\caption{For different latitudes between +65$^\circ$ to -65$^\circ$
longitudes from the central meridian, variation 
of different physical parameters 
(blue bar plot) such as area, radiative flux emitted by CH on the sun 
and at Earth and, apparent temperature 
structure of CH respectively. Red continuous line represents a least-square 
fit of the form $Y(\theta)=C_{0}+C_{1} sin^{2} \theta$ to different 
observed parameters (where $\theta$ is the latitude, $C_{0}$ and $C_{1}$ 
are constant coefficients determined from the least square fit). Whereas the 
red dashed lines represent the one standard deviation (which is computed 
from all the data points) error bands. $\chi^{2}$ is a measure of goodness of fit.
}
\end{center}
\end{figure}

\begin{figure}
\begin{center}
 \hskip 8ex 7(a) \hskip 40ex 7(b)
    \begin{tabular}{cc}
      {\includegraphics[width=18pc,height=18pc]{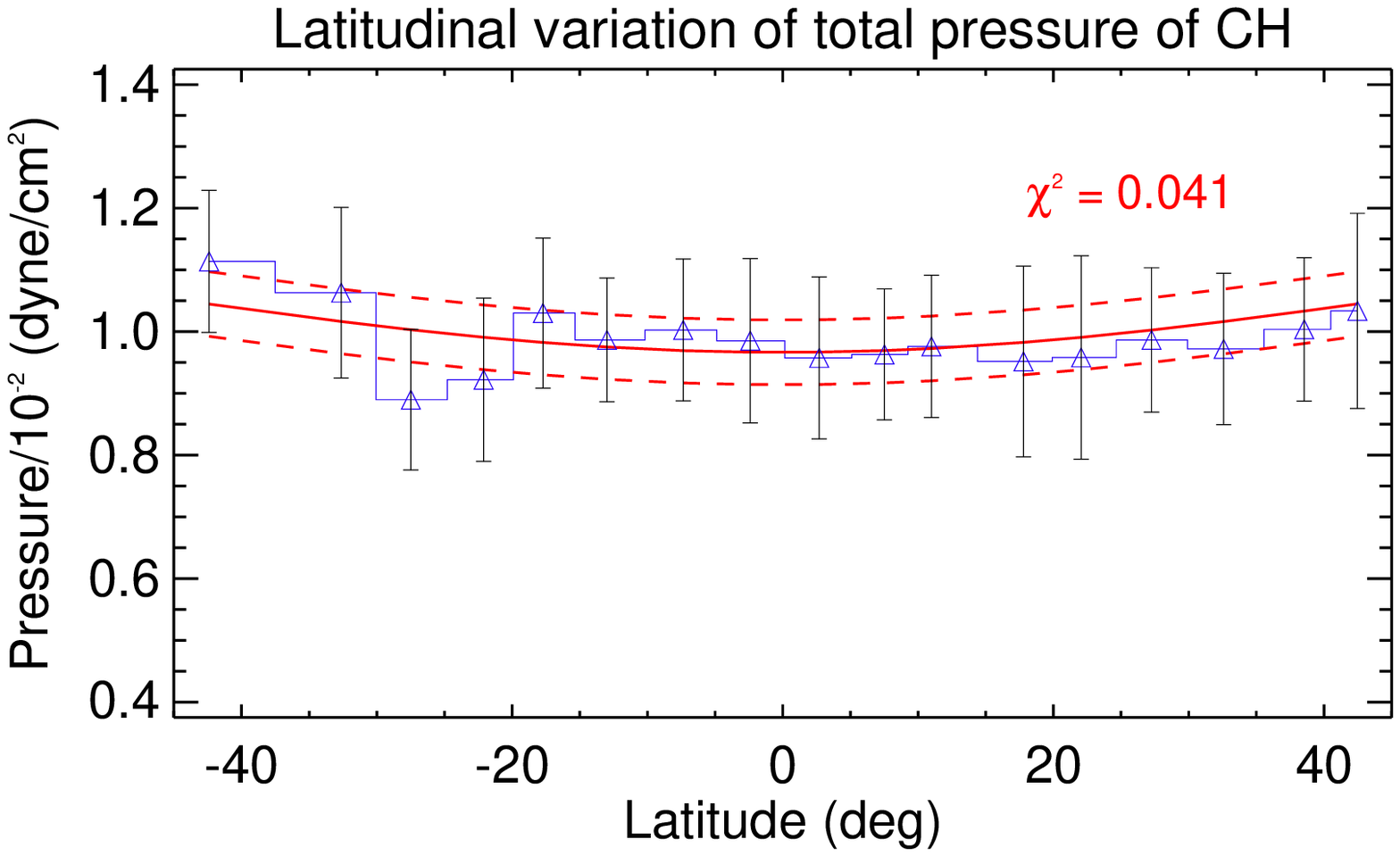}} &
      {\includegraphics[width=18pc,height=18pc]{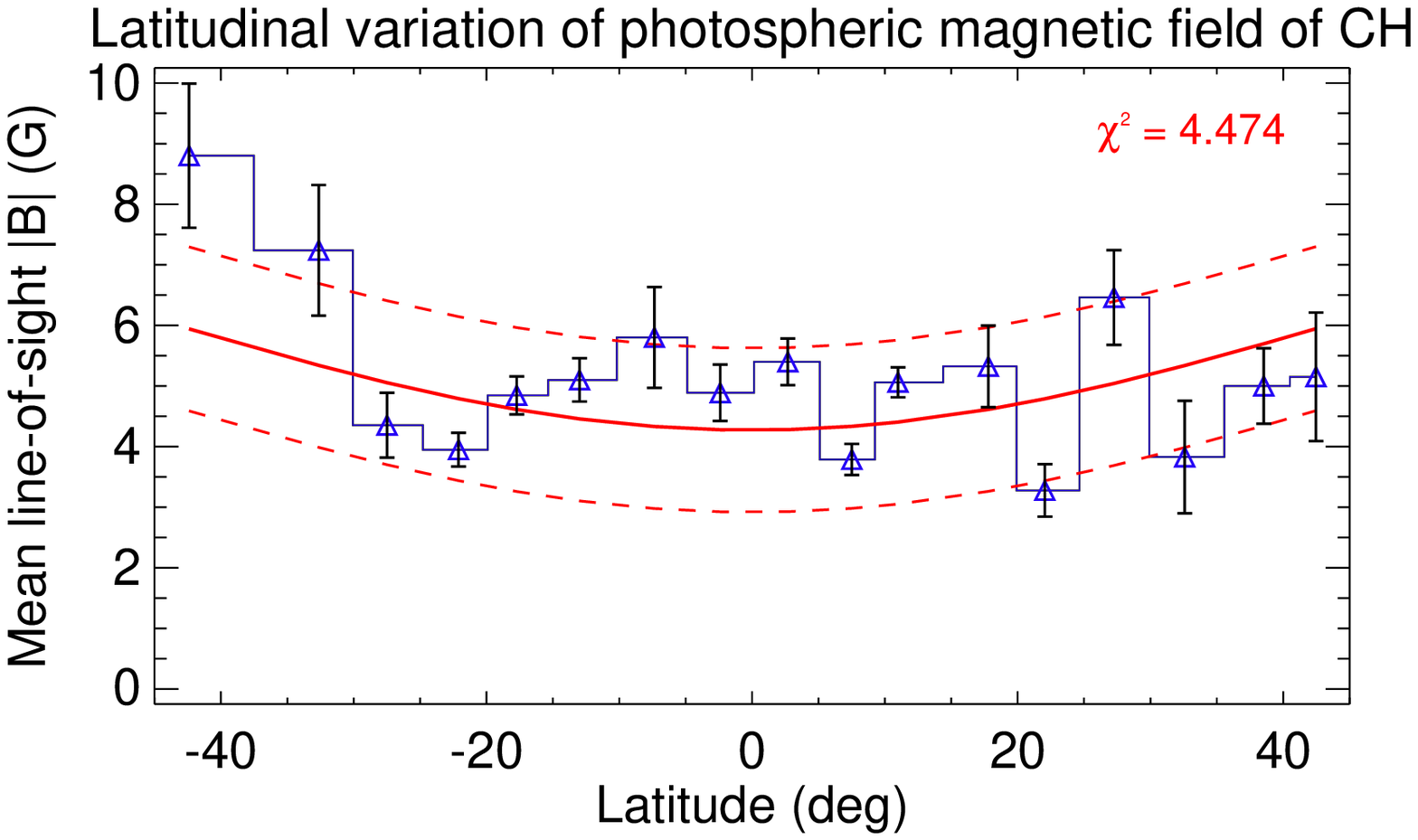}} \\
\end{tabular}

\begin{tabular}{cc}
      {\includegraphics[width=18pc,height=18pc]{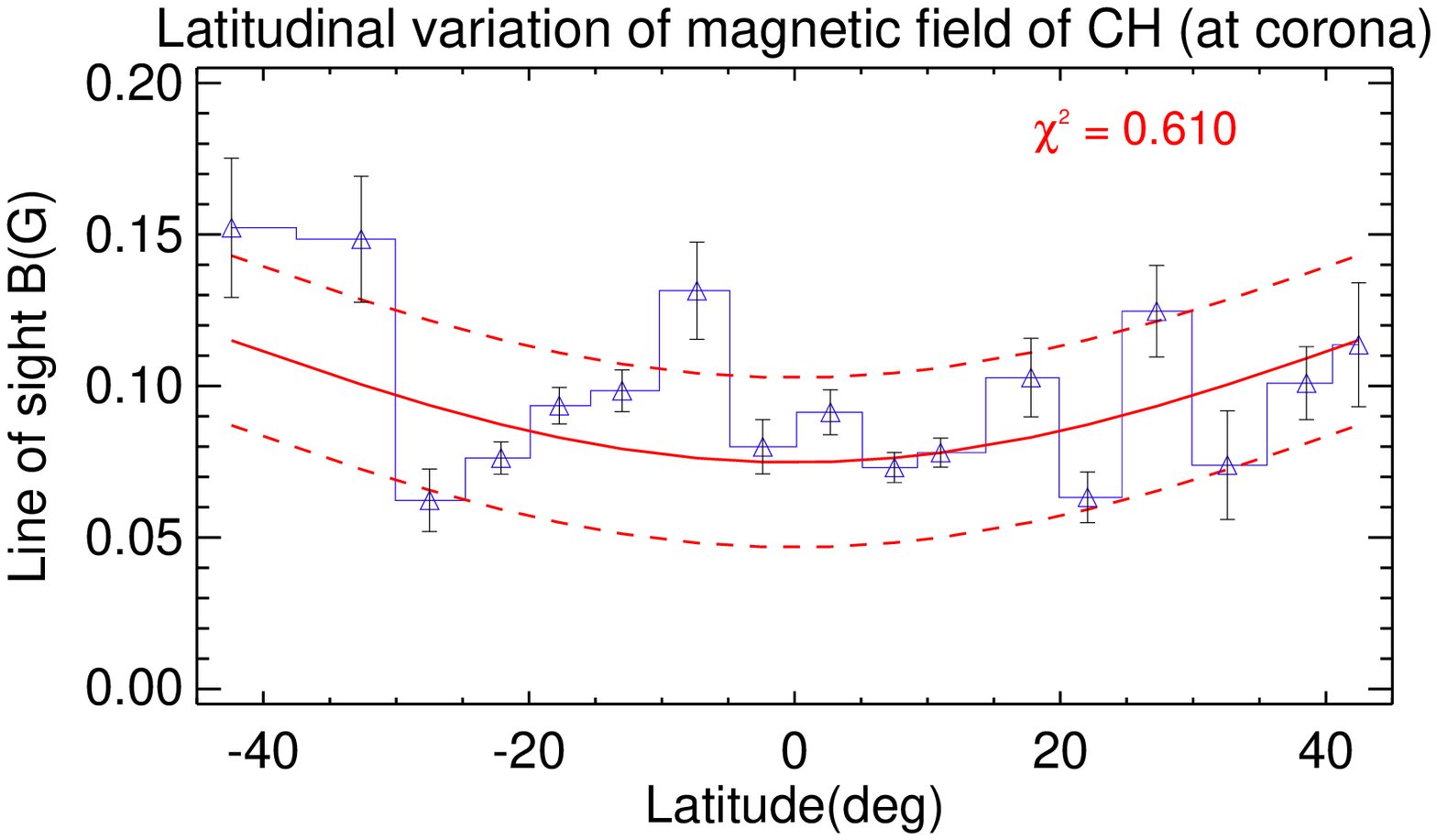}} &
{\includegraphics[width=18pc,height=18pc]{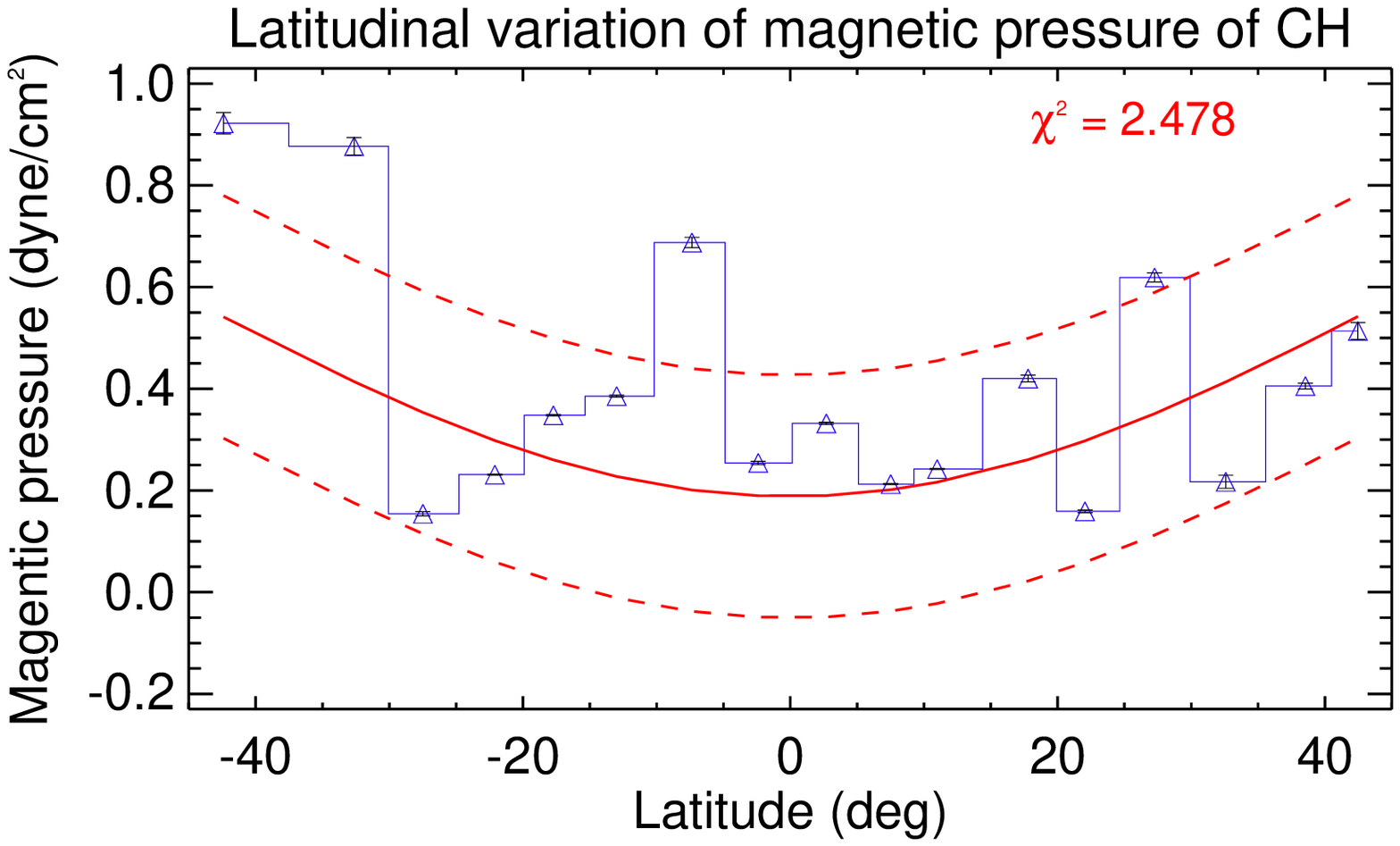}} \\
\end{tabular}
 \hskip 8ex 7(c) \hskip 40ex 7(d)
\caption{For different latitudes, between +65$^\circ$ to -65$^\circ$
longitudes from the central meridian, variation of different physical parameters
(blue bar plot) such as total pressure, magnitude of magnetic field structure
at the photosphere and at the corona and, magnetic pressure  of CH respectively. Red continuous line represents a least-square
fit of the form $Y(\theta)=C_{0}+C_{1} sin^{2} \theta$ to different
observed parameters (where $\theta$ is the latitude, $C_{0}$ and $C_{1}$
are constant coefficients determined from the least square fit). Whereas the
red dashed lines represent the one standard deviation (which is computed
from all the data points) error bands. $\chi^{2}$ is a measure of goodness of fit.
}
\end{center}
\end{figure}

\begin{figure}
\begin{center}
\hskip 8ex 8(a) \hskip 40ex 8(b)
    \begin{tabular}{cc}
      {\includegraphics[width=18pc,height=18pc]{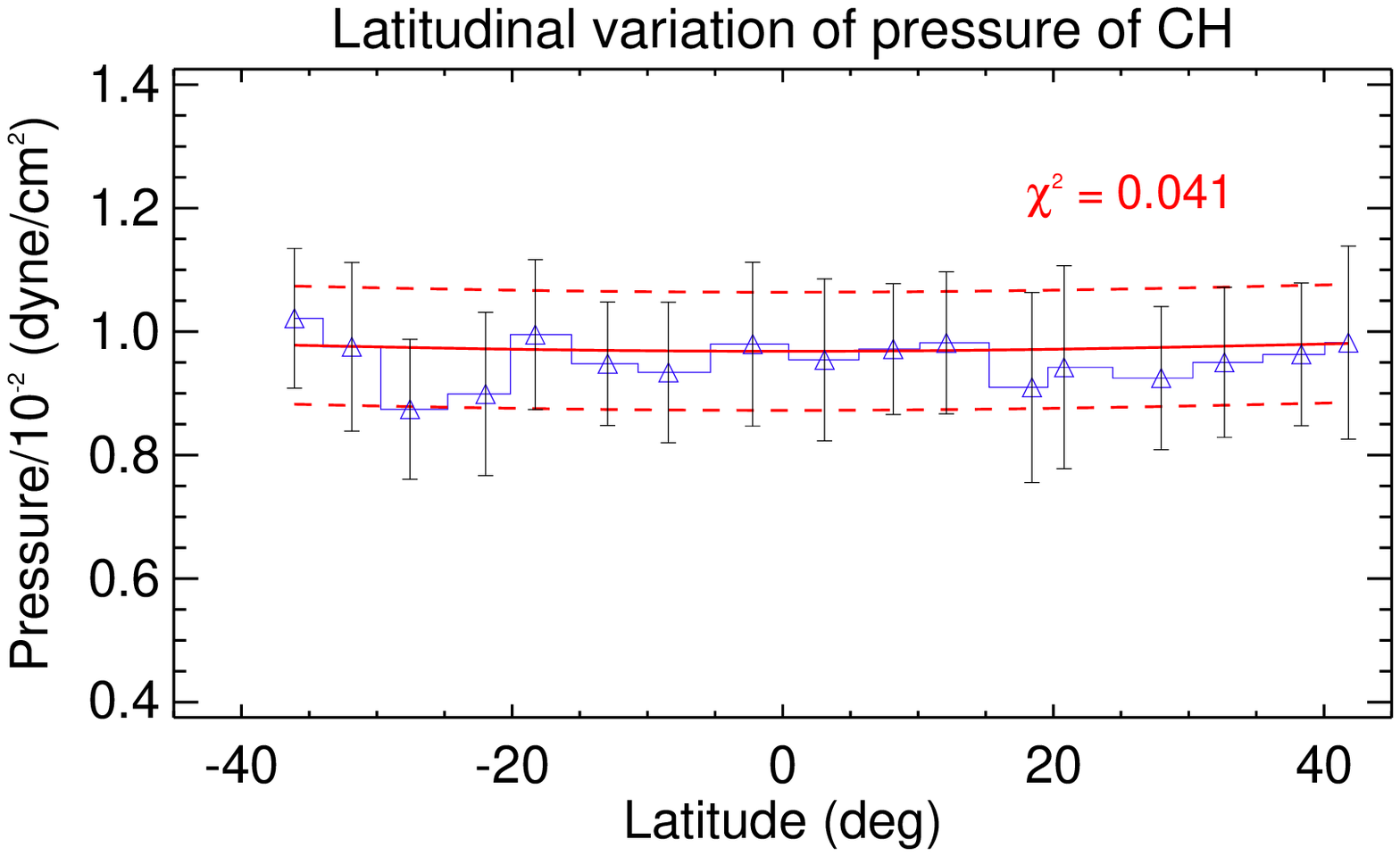}} &
      {\includegraphics[width=18pc,height=18pc]{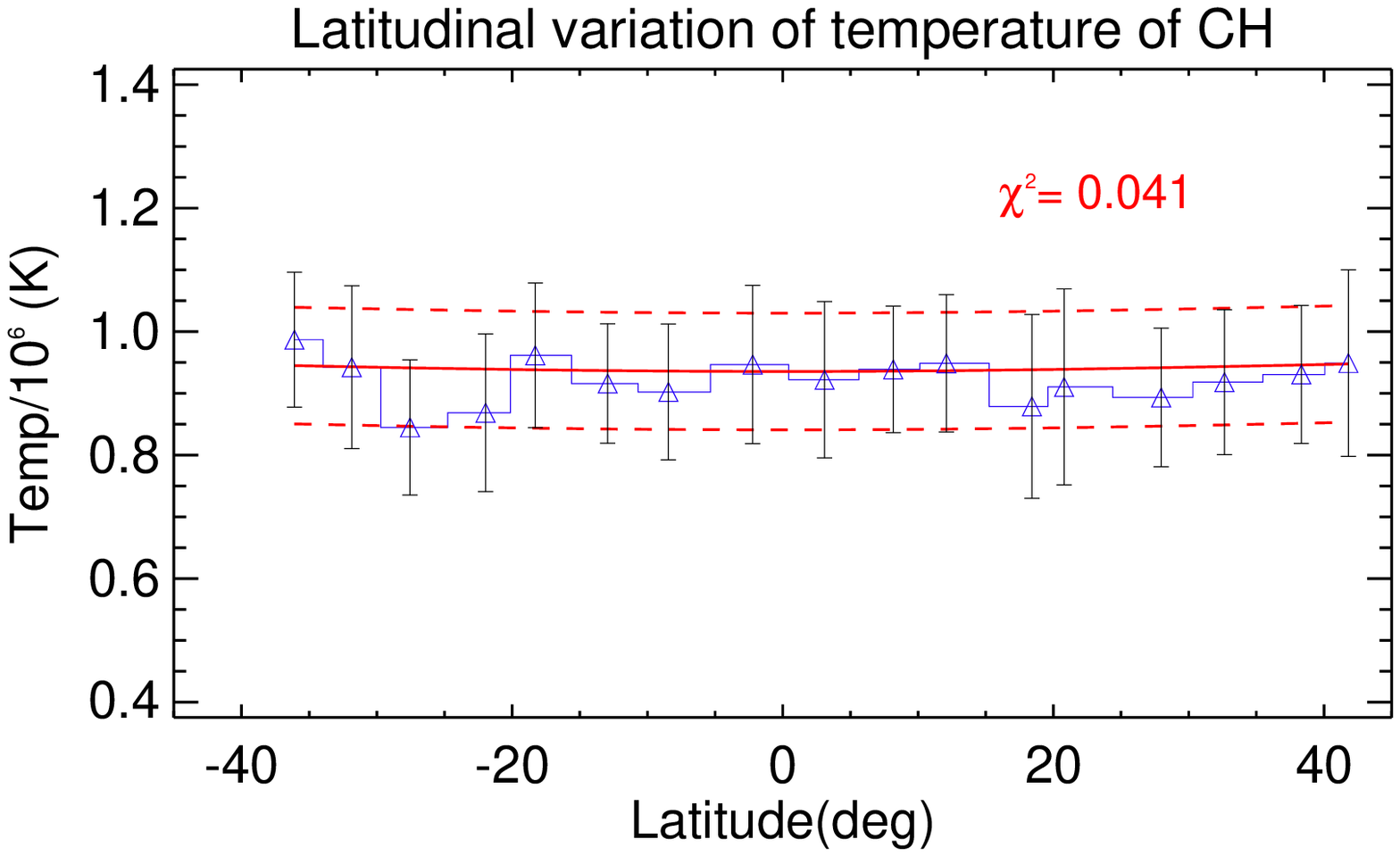}} \\
\end{tabular}
\caption{For different latitudes, between +65$^\circ$ to -65$^\circ$
longitudes from the central meridian, variation of thermal pressure and temperature
structure of CH. Red continuous line represents a least-square
fit of the form $Y(\theta)=C_{0}+C_{1} sin^{2} \theta$ to different
observed parameters (where $\theta$ is the latitude, $C_{0}$ and $C_{1}$
are constant coefficients determined from the least square fit). Whereas the
red dashed lines represent the one standard deviation (which is computed
from all the data points) error bands. $\chi^{2}$ is a measure of goodness of fit.
}
\end{center}
\end{figure}

\begin{figure}
\begin{center}
 \hskip 8ex 9(a) \hskip 40ex 9(b)
    \begin{tabular}{cc}
      {\includegraphics[width=18pc,height=18pc]{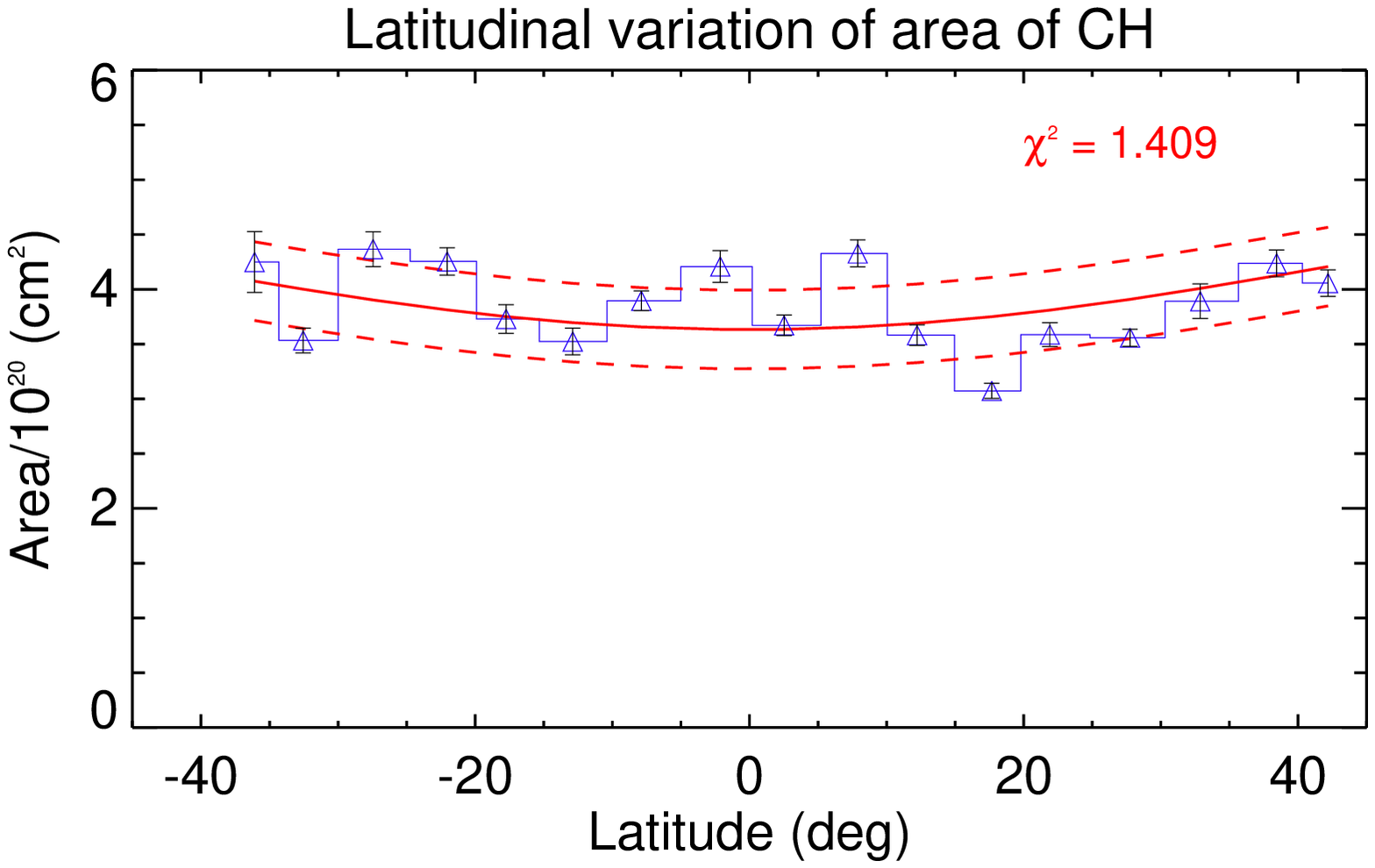}} &
      {\includegraphics[width=18pc,height=18pc]{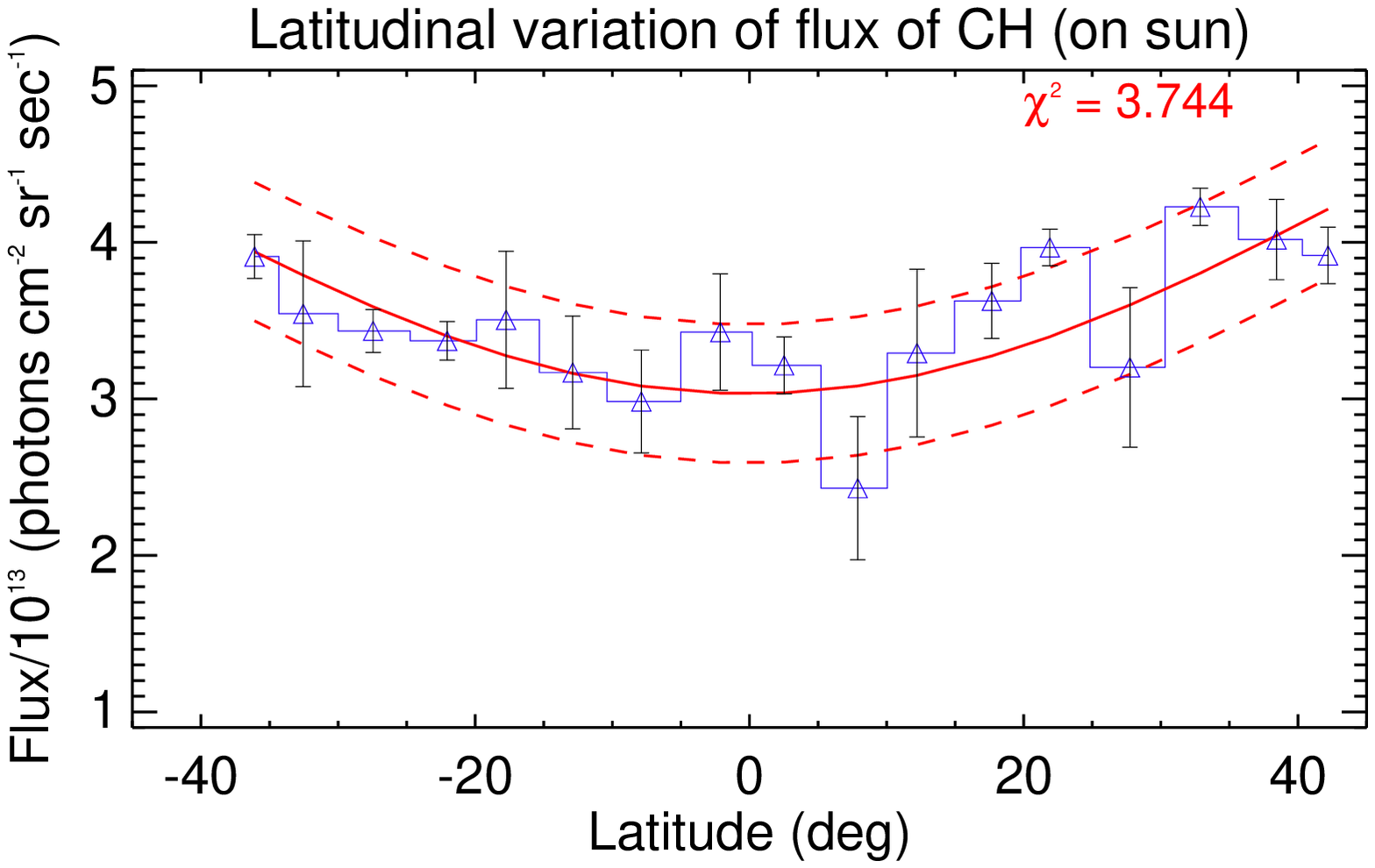}} \\
\end{tabular}

\begin{tabular}{cc}
      {\includegraphics[width=18pc,height=18pc]{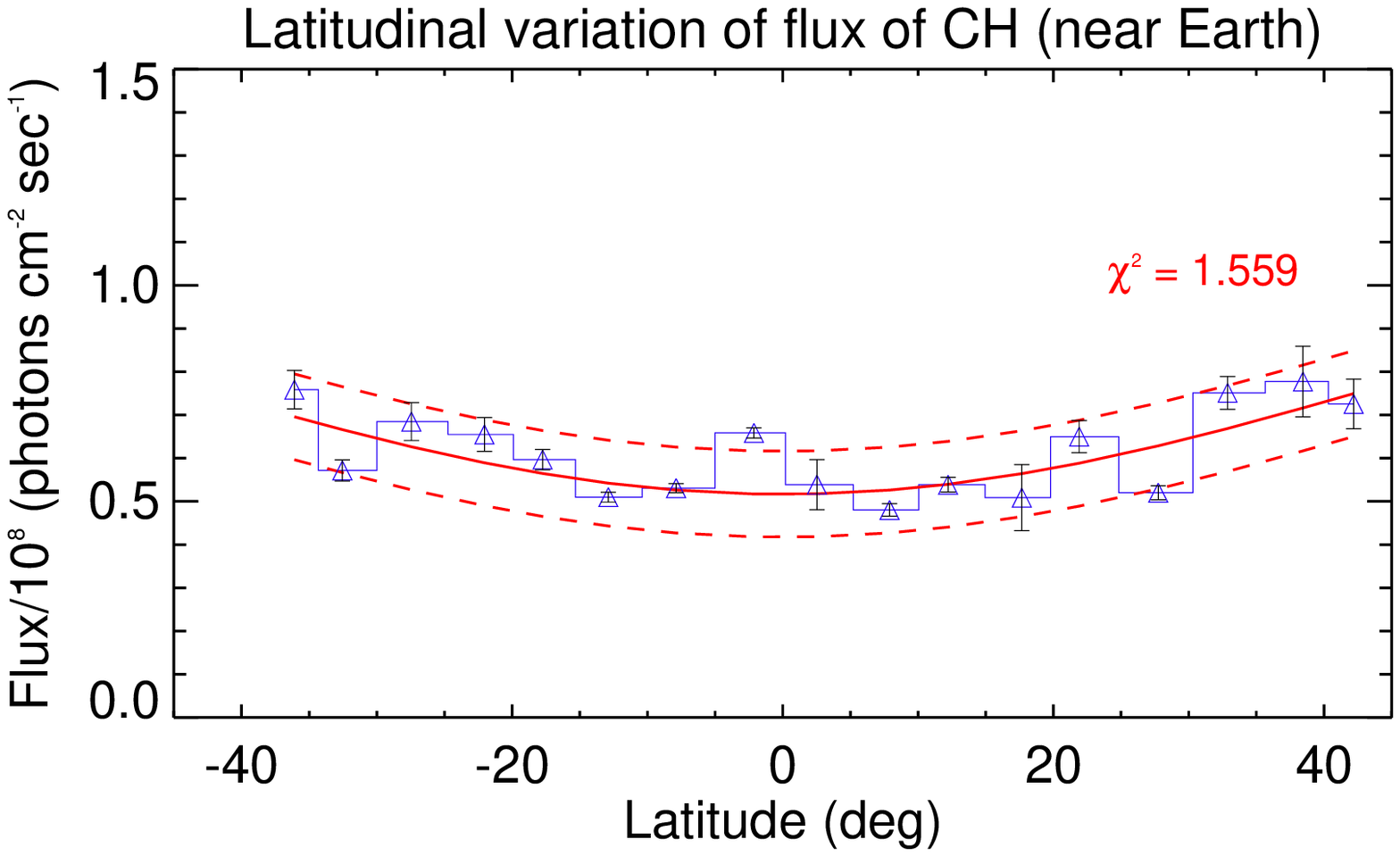}} &
      {\includegraphics[width=18pc,height=18pc]{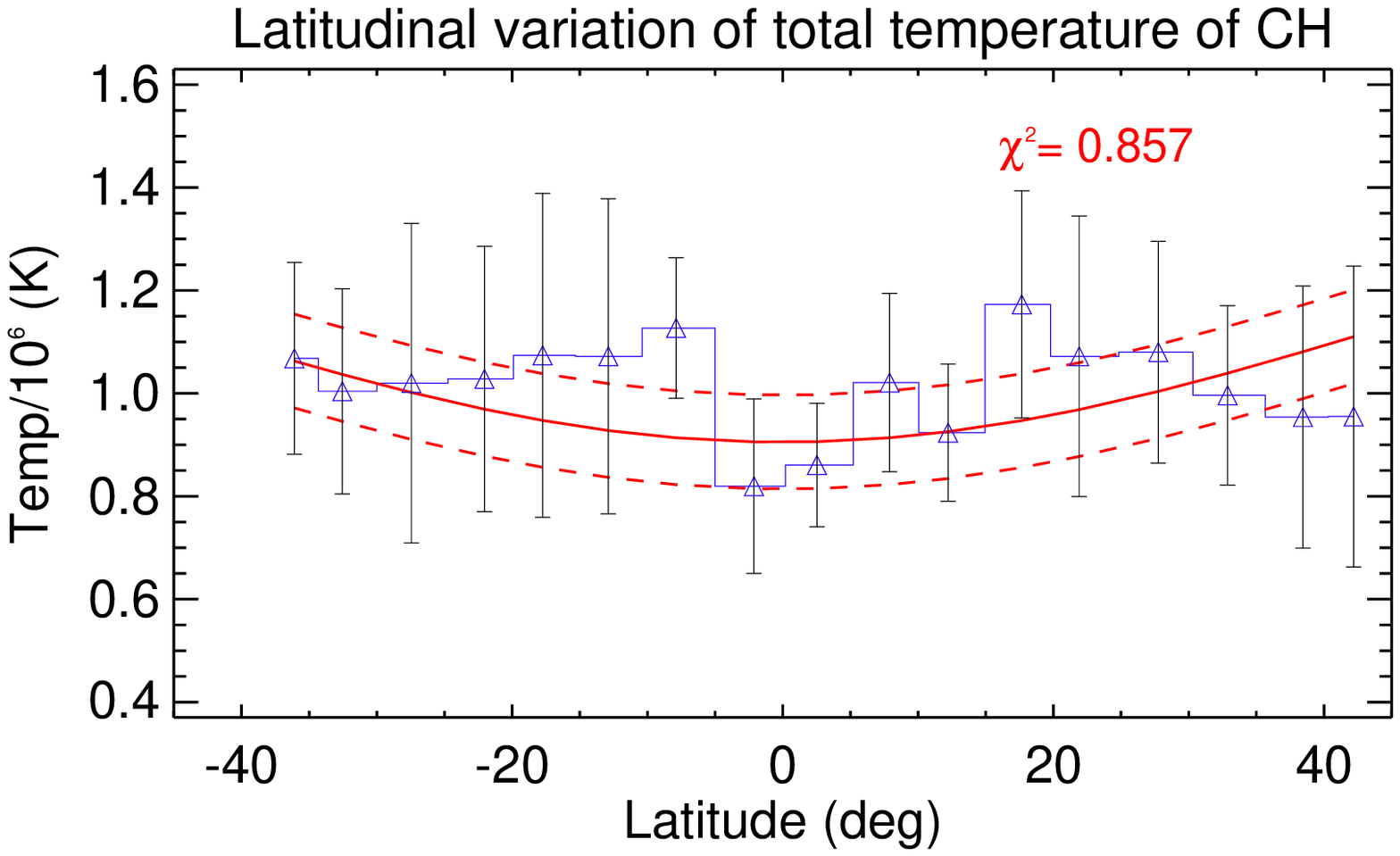}} \\
\end{tabular}
\hskip 8ex 9(c) \hskip 40ex 9(d)
\caption{ 
For different latitudes, between +45$^\circ$ to -45$^\circ$
longitudes from the central meridian, variation of different physical parameters
(blue bar plot) such as area, radiative flux emitted by CH on the sun
and near the Earth of CH respectively. Red continuous line represents a least-square
fit of the form $Y(\theta)=C_{0}+C_{1} sin^{2} \theta$ to different
observed parameters (where $\theta$ is the latitude, $C_{0}$ and $C_{1}$
are constant coefficients determined from the least square fit). Whereas the
red dashed lines represent the one standard deviation (which is computed
from all the data points) error bands. $\chi^{2}$ is a measure of goodness of fit.
}
\end{center}
\end{figure}

\begin{figure}
\begin{center}
 \hskip 8ex 10(a) \hskip 40ex 10(b)
    \begin{tabular}{cc}
      {\includegraphics[width=18pc,height=18pc]{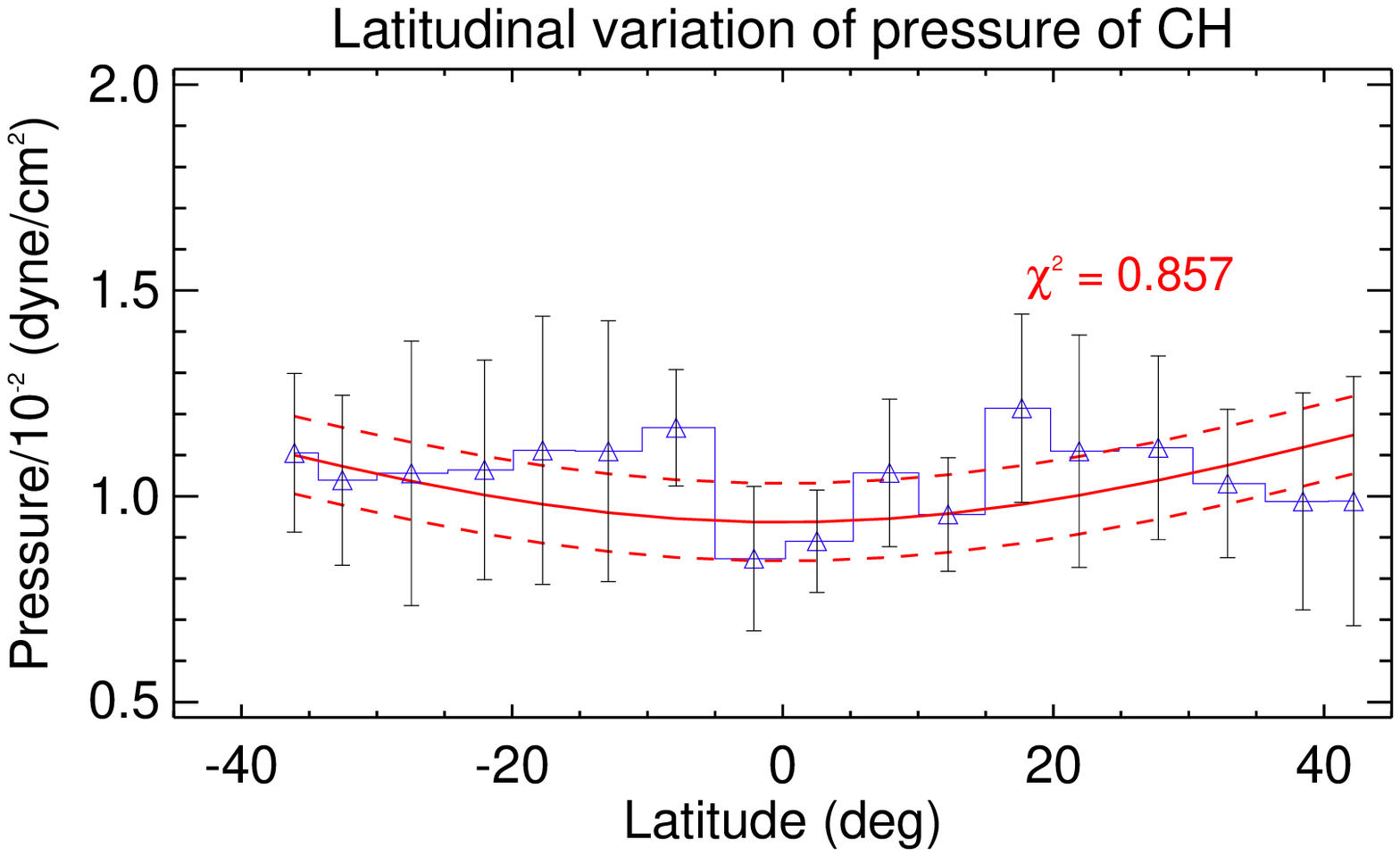}} &
      {\includegraphics[width=18pc,height=18pc]{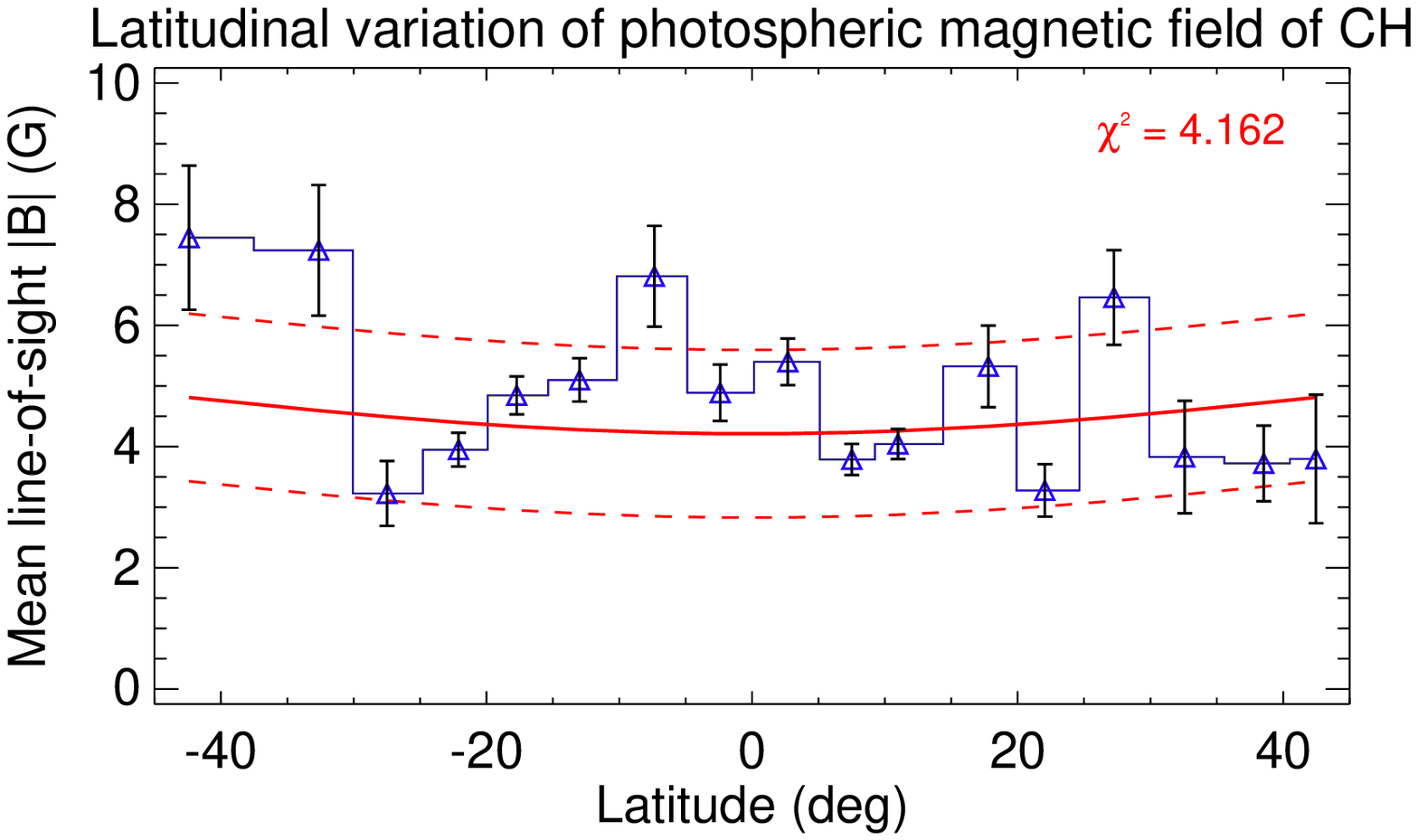}} \\
\end{tabular} 

\begin{tabular}{cc}
      {\includegraphics[width=18pc,height=18pc]{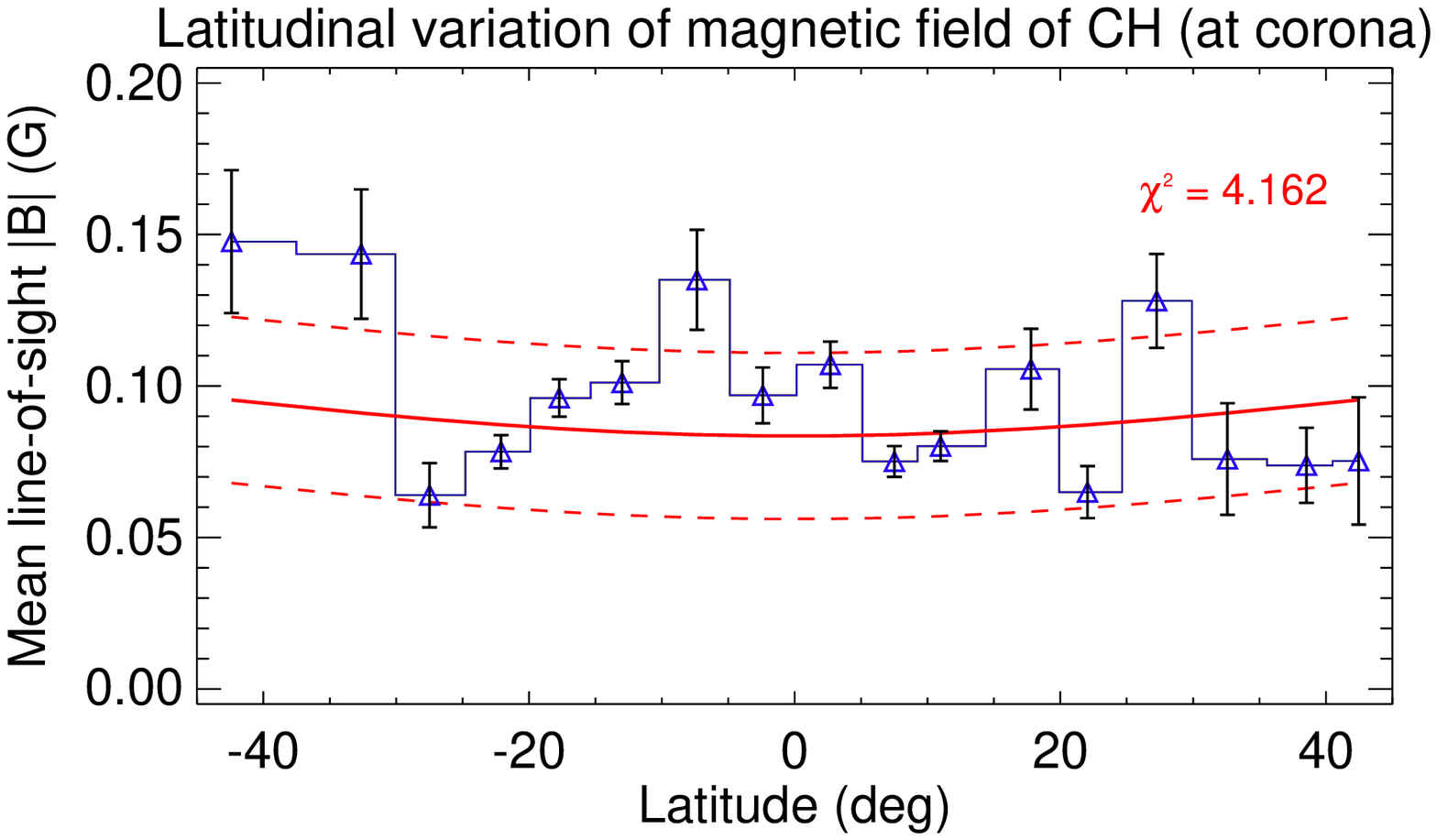}} &
      {\includegraphics[width=18pc,height=18pc]{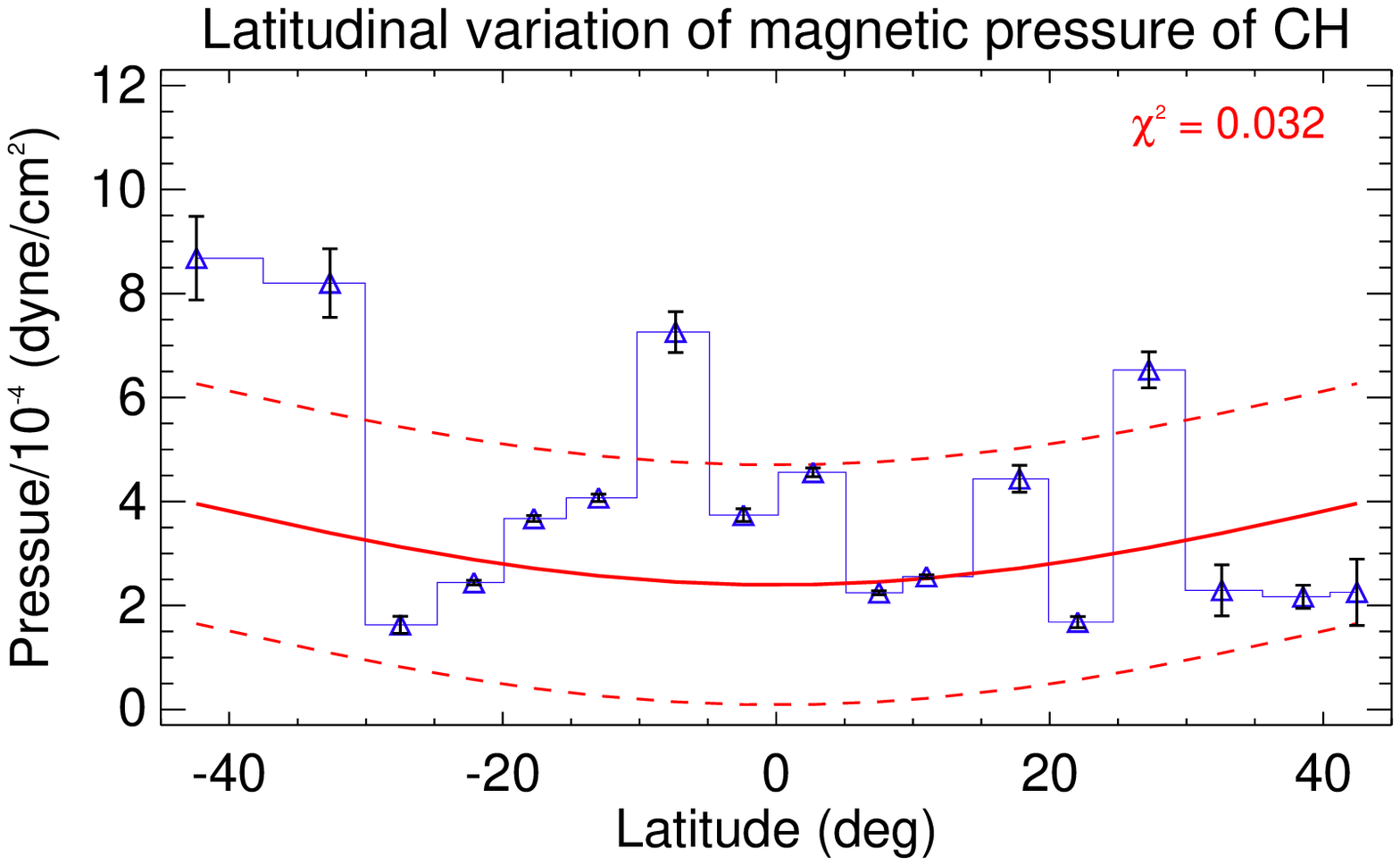}} \\
\end{tabular} 
 \hskip 8ex 10(c) \hskip 40ex 10(d)
\caption{ 
For different latitudes, between +45$^\circ$ to -45$^\circ$
longitudes from the central meridian, variation of different physical parameters
(blue bar plot) such as total pressure, magnitude of magnetic field structure
at the photosphere and at the corona and, magnetic pressure of CH respectively. 
Red continuous line represents a least-square
fit of the form $Y(\theta)=C_{0}+C_{1} sin^{2} \theta$ to different
observed parameters (where $\theta$ is the latitude, $C_{0}$ and $C_{1}$
are constant coefficients determined from the least square fit). Whereas the
red dashed lines represent the one standard deviation (which is computed
from all the data points) error bands. $\chi^{2}$ is a measure of goodness of fit.
}
\end{center}
\end{figure}

\begin{figure}
\begin{center}
    \hskip 8ex 11(a) \hskip 40ex 11(b)
    \begin{tabular}{cc}
      {\includegraphics[width=18pc,height=18pc]{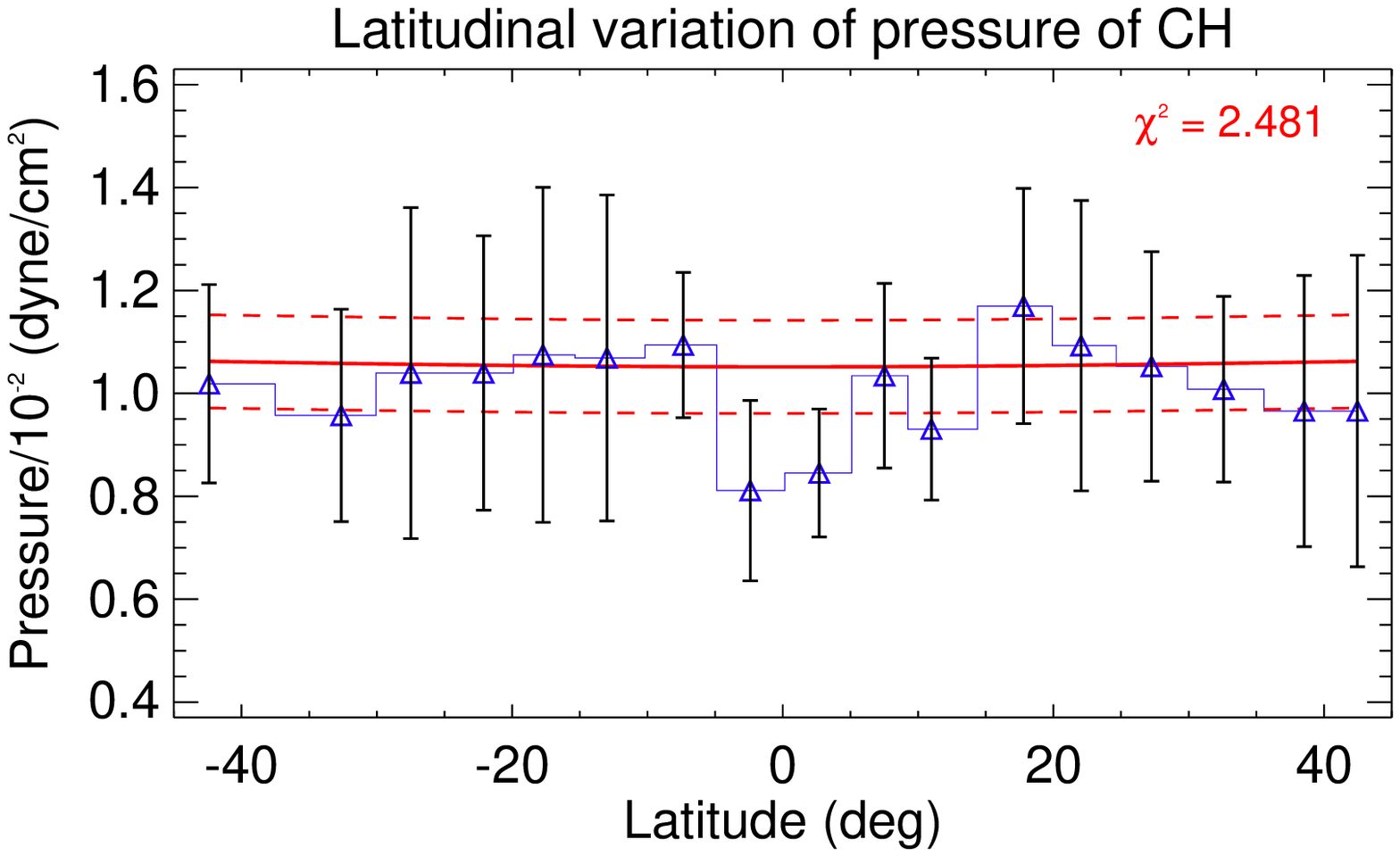}} &
      {\includegraphics[width=18pc,height=18pc]{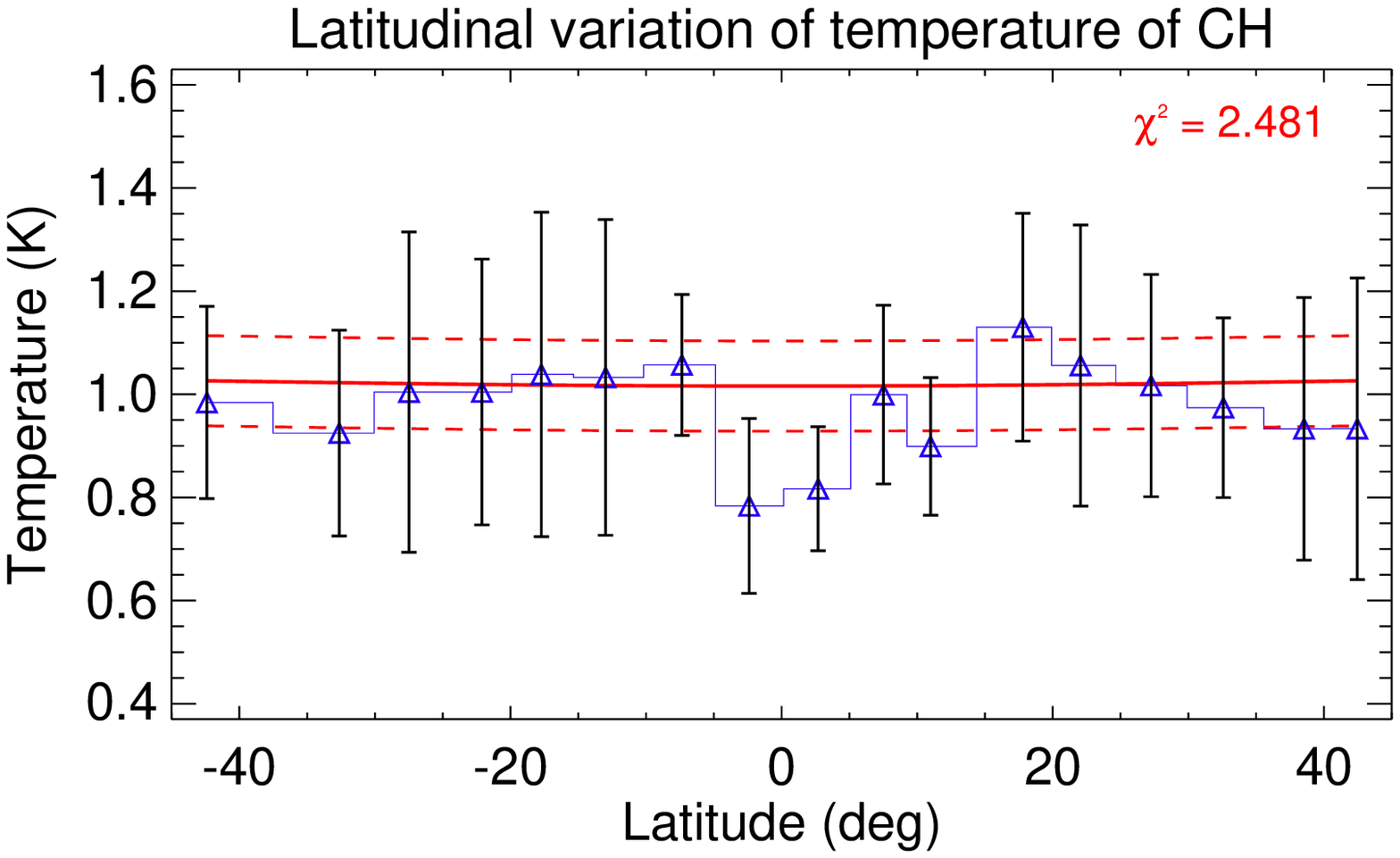}} \\
\end{tabular}
\caption{
For different latitudes, between +45$^\circ$ to -45$^\circ$
longitudes from the central meridian, variation of thermal pressure and temperature
structure of CH. Red continuous line represents a least-square
fit of the form $Y(\theta)=C_{0}+C_{1} sin^{2} \theta$ to different
observed parameters (where $\theta$ is the latitude, $C_{0}$ and $C_{1}$
are constant coefficients determined from the least square fit). Whereas the
red dashed lines represent the one standard deviation (which is computed
from all the data points) error bands. $\chi^{2}$ is a measure of goodness of fit.
}
\end{center}
\end{figure}

\section {Discussion and conclusions}
There are two very interesting results that need worthy to be discussed here.
First result is the magnitude of magnetic field structure of coronal holes
that increases from equator to both poles of the
sun.  One can argue from the 
previous studies (Harvey et al 1982; Webb and Davis 1985;
 Abramenko et.al 2009) that this result could be due
to occurrence of coronal holes during the evolution of 
solar cycle as the polar coronal 
holes have stronger magnetic fields than the coronal holes at low latitudes.
Although inferred result of latitudinal variation of strength
of magnetic field structure of coronal holes from the present
study is consistent with the
results of previous studies, question remains why high latitude 
coronal holes occur with high average
magnetic field strengths than the low latitude coronal holes.

This important observed and inferred information
of latitudinal variation of strength of magnetic field
 probably suggests coronal holes' origin that is
not consistently understood. As the coronal holes
 are unipolar magnetic field structures,
their origin can be understood from the global nature of
magnetic field structure of the sun. Probably, we believe, 
following conjecture on genesis of coronal holes is consistent
with the result of latitudinal dependency of magnitude
of magnetic field structure of the coronal holes.

During the period of minimum
solar activity, white light pictures taken during total solar eclipse,
one can notice the dipole like magnetic field structure delineated along
the intensity patterns (rays) that are originated from the
two poles. Infact observational (Stenflo 1993)  and theoretical
studies (Hiremath and Gokhale 1995 and, references
there in) can not rule out such a large-scale global dipole like
magnetic field structure,  may be of primordial
origin. 
 Offcourse we have to make a clear distinction between
the ``steady" and ``time dependent" parts of solar magnetic field
structure. ``Steady" part of solar magnetic field structure 
has a time scale of billion of years (Hiremath and Gokhale 1995
and reference therein). Whereas time dependent part
of solar magnetic field structure has a 22 yrs time
scale. For understanding genesis
of coronal hole, in the following conjecture, we
invoke the large-scale ``steady" part of the magnetic 
field structure in order to be compatible with the
inferred latitudinal variation of magnetic field structure of the coronal
hole.

Observational (Stenflo 1993) estimates yield
the intensity to be $\sim$ 1 G. Whereas, for matching of 22 year
of magnetic periodicity, previous study (Hiremath 
and Gokhale 1995 and, references
there in) estimates sun's global dipole like 
magnetic field structure to be $\sim$ 0.01 G. 
This result is also consistent with the recently 
estimated average magnetic field structure of the $\sim$ 0.028 G
(Kotov 2015).
Although genesis of coronal holes is
debatable, for the consistency of inferred coronal
hole magnetic field structure whose intensity increases
from equator to pole, we present probable mechanism of CH formation as
follows.

If the large-scale coronal poloidal magnetic field structure is
perturbed, Alfven waves are produced whose interference
pattern leads to formation of coronal hole structure. If $B_{p}$
is intensity of large-scale steady part of magnetic field structure, then
resulting amplitude $\delta B_{p}$ of Alfven waves is same order 
as that of intensity of original magnetic field structure. That is
$\delta B_{p}$ is $\sim$ $B_{p}$. Although steady part of poloidal 
part of magnetic field
structure probably consists of combined (uniform, dipole and quadrupole)
filed structure (Hiremath and Gokhale 1995), for
the sake of simplicity, let us consider dipole like field structure
only. For a particular radius, intensity or magnitude of magnetic
field structure varies as $sin^{2} \lambda$ (where $\lambda$ is
observed latitude; here $\lambda$=0$^\circ$ is equator and 
$\lambda$=90$^\circ$ is pole) whose magnitude 
increases from equator to pole. That means,
as presented in Figures (7(c) and 10(c)), 
coronal holes originated at the equator
must have less magnitude of magnetic field structure compared
to the coronal holes that originate near the poles. If one accepts
a naive concept that coronal holes are formed due to Alfven wave
perturbations of poloidal component of magnetic field
structure, as the amplitude of magnetic field structure of Alfven
waves is at least same order as that of steady part of magnetic field structure,
hence it is not surprising that average (see the fits) 
strength of magnetic field
structure is same as strength (in the range of 0.01-1 G) of
steady magnetic field structure as estimated by observational (Stenflo 1993;
 Kotov 2015) and theoretical (Hiremath and Gokhale 1995) studies. 
 It is to be noted that this speculative explanation for the
origin of coronal holes has to be treated cautiously unless
some other studies also agree with our conjecture that
coronal holes originate from the Alfven perturbations of
global large-scale magnetic field structure.

Another interesting result is that variation of thermal structure
(especially actual temperature) of
coronal holes is independent of latitudes. As coronal hole is
a magnetic flux tube, for the steady state of magnetic field
structure and no gain of magnetic flux during coronal
hole evolution, condition of infinite conductivity
leads to isorotation of coronal holes with the surrounding 
ambient plasma rotation.
That means coronal hole magnetic flux bundle follows the path of
isorotational contours. To be precise, coronal holes that
might have formed due to Alfven wave perturbations travel along
the large scale magnetic field structure parallel
to isorotational contours. Infact, in the previous study
(Hiremath and Gokhale 1995), we have shown that
large-scale magnetic field structure, may be of primordial origin,
consists of combined magnetic field structure
in the radiative core and current free (combination
of dipole and quadrupole like field structures that are embedded
in the uniform) field structure in the
convective envelope and both the structures in turn isorotate with
the internal solar plasma rotation as inferred by the helioseismology.

From the above discussion, it is clear  that rotation rate
of coronal holes and rotation rate of ambient solar plasma depend
upon each other. Helioseismic inferences (Hiremath 2013, 2016 and
references there in) yield rigid body rotation
in the radiative core and differential rotation in the convective
envelope. That means if coronal holes originate only in the
convective envelope, they must rotate differentially. Otherwise coronal holes
likely to rotate rigidly if coronal holes originate in the radiative
core. On the theoretical (Gilman 1977; Golub {\em et. al.} 1981; Jones 2005) and observational (Hiremath and Hegde 2013)
inferences,  it is argued that coronal holes
probably originate below the convective envelope which in turn 
implies that coronal holes rotate rigidly (as the
radiative core rotates rigidly).  On other hand, by evolving magnetic
diffusion equation, Wang and Sheeley (1993) simulate the
rotation rate of the coronal holes from the current free nature
of the coronal field that is rotating with the distorted
active region fields.

If one considers curl of momentum equation
in the cylindrical coordinates and for steady angular velocity gradient, angular
velocity of solar plasma is balanced (Chandrasekhar 1956; 
Brun, Antia and Chitre 2010 and references there in) by the combined forces
due to stretching/tilting of vorticity due
to velocity gradients, advection of vorticity by the flows,
turbulent and Reynold stresses, Maxwell stresses, baroclinic forces, etc.
If one assumes other forces are negligible,
then gradient of angular velocity is balanced by the baroclinic forces only.
In cylindrical coordinates, the relationship can be expressed in the form of equation
${{\partial\Omega }\over{\partial z}}={g\over rc_{p}} {{\partial <S^{'}>}\over{\partial z}}$,
where $\Omega$ is angular velocity of the solar plasma, $g$ is acceleration due to gravity,
$c_{p}$ is specific heat at the constant pressure, 
${g\over c_{p}}$ is the adiabatic temperature gradient, $r$ is the radial 
variation and $<S^{'}>$ is entropy of the ambient medium. This equation
is called thermal wind balance equation (Brun, Antia and Chitre 2010). That means,
unless there is a temperature difference between the
pole and equator, angular velocity of the solar plasma
can not be differential with respect to latitude
and also can not be maintained. As the coronal holes isorotate
with the solar plasma, this equation also implies that
unless there is a temperature structure that
varies from equator to pole, coronal holes can not
rotate rigidly. However, present study yields the temperature
structure (and hence entropy) of coronal holes independent of solar latitude.
 Hence, above thermal wind balance equation implies that ${\partial\Omega }\over{\partial z}$=0.
That means coronal holes must rotate rigidly or rotation
rate of coronal holes is independent of solar latitude.
Infact this reasoning also matches with the recent results (Hiremath and Hegde 2013)
of rigid body rotation rates of coronal holes derived
from the SOHO 195 $\AA$ data.

To conclude this study, for the years 2001-2008,  near equatorial
 coronal holes detected 
from the SOHO/EIT images are used to understand the area evolution
and, latitudinal variation of thermal and magnetic field structure. 
Different estimated physical parameters
of the coronal holes are: area $\sim$ $3.8(\pm 0.5) \times 
10^{20}$ $cm^{2}$,
radiative flux at the sun $\sim$ $2.3(\pm0.2) \times 
10^{13}$ photons $cm^{-2}$
 $sec^{-1}$, radiative energy $\sim$ $2.32 (\pm 0.5) \times 
10^{3}$ ergs $cm^{-2} sec^{-1}$, 
temperature structure $\sim$ $0.94 (\pm 0.1) \times 10^{6}$ K
and magnitude of magnetic field structure estimated to be 
$\sim$ $0.08\pm 0.02$ G.
  
\centerline{\bf Acknowledgments}
\vskip 0.2cm 
This work has been carried out under  ``CAWSES India Phase-II
program of Theme 1'' sponsored by Indian Space Research Organization(ISRO),
Government of India. SOHO is a mission of international cooperation between ESA and NASA.
\vskip 0.5cm
\centerline {\bf REFERENCES}
\vskip 0.3cm
\noindent Abramenko, V., Yurchyshyn, V \& Watanabe, H. 2009, \solphys, 260, 43 

\noindent Aschwanden, M, J., 2004, Physics of the Solar Corona - An Introduction,  Praxis Publishing Ltd., Chichester, UK, and 
Springer-Verlag Berlin 

\noindent Bauer, S. J., 1973, in Physics of Planetary Ionospheres, Springer-Verlag, p. 47

\noindent Bohlin, J. D. 1977, \solphys, 51, 377

\noindent Brun, A. S., Antia, H. M. \& Chitre, S. M., 2010, \aap, 510, 33

\noindent Cally, P. S. 1986, \solphys, 103, 277

\noindent Cally, P. S. 1987, \solphys, 108, 183-

\noindent Chandrasekhar S.,1956, \apj, 124, 231

\noindent Chiuderi, D. F., Avignon, Y \& Thomas, R. J., 1977, \solphys, 51, 143 

\noindent Chiuderi, D. F., Landi, E., Fludra, A. \& Kerdraon, A., 1999, \aap, 348, 261

\noindent Choi, Y., Moon, Y.-J., Choi, S., Baek, J.-H., Kim, S. S., Cho, K.-S., \& Choe, G. S. 2009, \solphys, 254, 311

\noindent Cranmer, S.R. 2009, Living Reviews in \solphys, 6, 3

\noindent David, C., Gabriel, A. H., Bely-Dubau, F., Fludra, A., Lemaire, P. \& Wilhelm, K. 1998, \aap, 336, L90

\noindent Davila, J. M. 1985, \apj, 291, 328

\noindent Delaboudiniere, J. P. et al. 1995, \solphys, 162, 291

\noindent Doschek, G. A. \& Feldman, U. 1977, \apj, 212, L143

\noindent Doschek, G. A., Warren, H. P., Laming, J. M., Mariska, J. T., Wilhelm, K., Lemaire, P.,
Schuehle, U., \& Moran, T. G. 1997, \apj letters, 482, 109

\noindent Doschek, G. A. \&  Laming, J. M. 2000, \apj, 539, L71

\noindent Doyle, J. G., Chapman, S., Bryans, P., Pérez-Suárez, D., 
Singh, A., Summers, H. \& Savin, D. W. Research in Astronomy and Astrophysics, 10, 91, 2010

\noindent Dudok de Wita, T \& Watermanna, J., 2010, Comptes Rendus Geoscience
342, 259

\noindent Dwivedi, B. N \& Mohan, A., 1995, \solphys, 156, 197

\noindent Dwivedi, B. N., Mohan, A \& Wilhelm, K. 2000, Advances in Space Research, 25, 1751

\noindent Esser, R., Fineschi, S., Dobrzycka, D., Habbal, S. R., Edgar, R. J., 
Raymond, J. C., Kohl, J. L. \& Guhathakurta, M. 1999, \apj, 510, L63

\noindent Fathy, I., Amory-Mazaudier, C., Fathy, A., Mahrous, A. M., 
Yumoto, K. \& Ghamry, E. 2014, Journal of Geophysical Research: Space Physics, 
119, 4120

\noindent Fla, T., Habbal, S. R., Holzer, T. E \& Leer, E. 1984, \apj, 280, 382

\noindent Gilman, P. A. 1977, Coronal holes and high speed wind streams Conference, ed. Zirker, J. B. 331

\noindent Golub, L., Rosner, R., Vaiana, G. S., \& Weiss, N. O. 1981, ApJ, 243, 309

\noindent Guennou, C.; Auchère, F.; Soubrié, E.; Bocchialini, K.; Parenti, S \& Barbey, N. 2012, \apjs, 203, 26, 14

\noindent Habbal, S. R., Esser, R., \& Arndt, M. B. 1993, \solphys, 413, 435

\noindent Habbal, S. R. 1996, Solar and Interplanetary Transients, proceedings of IAU Colloquium 154, edited by S. Ananthakrishnan; A. Pramesh Rao., p. 49

\noindent Hahn, M., Landi, E., \& Savin, D. W. 2011, \apj, 736, 101

\noindent Hara, H., Tsuneta, S., Acton, L. W., Bruner, M. E., Lemen, J. R. \& Ogawara, Y. 1994, PASJ, 46, 493

\noindent Hara, H., Tsuneta, S., Acton, L. W., Bruner, M. E., Lemen, J. R. \& Ogawara, Y. 1996, Advanced Space Research, 17, 4

\noindent Harvey, J. W. \& Sheeley, N. R., Jr. 1979, \ssr, 23, 139

\noindent Harvey, K. L., Harvey, J. W., \& Sheeley, N. R., Jr. 1982, \solphys, 79, 149

\noindent Hegde, M., Hiremath, K. M., Doddamani, V. H., Gurumath, S. R. 2015, 
Journal of Astrophysics and Astronomy, 36, 355

\noindent Hegde, M., Hiremath, K. M., Doddamani, V. H. 2014, Advances in Space 
Research, 54, 272 

\noindent Hiremath, K. M. 1994, Study of Sun's Long Period Oscillations, 
Ph.D Thesis, Bangalore University, 1994, p. 142, http://prints.iiap.res.in/handle/2248/122

\noindent Hiremath, K. M \& Gokhale, M. H. 1995, \apj, 448, p.437

\noindent Hiremath, K. M. 2001, Bulletin of the Astronomical Society of India,
 29, 169

\noindent Hiremath, K. M \& Mandi, P. I. 2004, New Astronomy, 9, 651

\noindent Hiremath, K. M. 2009, Sun and Geosphere, 4, 16

\noindent Hiremath, K. M. 2013, in Seismology of the Sun: Inference of 
Thermal, Dynamic and Magnetic Field Structures of the Interior, 2013, 
p. 333, springer

\noindent Hiremath, K. M., \& Hegde, M. 2013, \apj, 763, 137

\noindent Hiremath, K. M., Hegde, M., \& Soon, W. 2015, New Astronomy, 35, 8

\noindent Hiremath, K. M. 2016, in Cartography of the Sun and the Stars., 
Eds: Rozelot, Jean-Pierre, Neiner, Coralie, p. 85

\noindent Hegde, M., Hiremath, K. M., Doddamani, V. H. \&
Gurumath, S. R. 2015, Journal of Astrophys and Astronomy, 36, 355

\noindent Hinteregger, H. E. 1976, Journal of Atmospheric and Terrestrial Physics, 38, 791

\noindent Insley, J.E., Moore, V., \& Harrison, R.A. 1995, \solphys, 160, 1

\noindent Japaridze, D. R., Bagashvili, S. R., Shergelasvili, B. M.,
 Chargeishvili, B. B. 2015, Astrophysics, 58, 575

\noindent Jones, H. P. 2005, Large-scale Structures and their Role in Solar Activity ASP Conference, eds, Sankarasubramanian, K., Penn, M., \& Pevtsov, A. 346, 229

\noindent Kahler, S. W \& Hudson, H. S. 2001, \jgr, 106, A12, 29239

\noindent Karachik, N. V \& Pevtsov, A, A. 2011, \apj, 735, 6

\noindent Kotov, V. A., 2015, Advances in Space Research, 55,979

\noindent Kretzschmar, M., Dudok de Wita, T., Lilensten, J., Hochedez, J.-F., Aboudarham, J., Amblard, P.-O., Auchère, F, \& Moussaoui, S. 2009, Acta Geophysica, 57, 42

\noindent Krieger, A.S., Timothy, A.F., \& Roelof, E.C. 1973, \solphys, 29, 505

\noindent Krista, L. D. 2011, Ph.D. Thesis, University of Dublin, Trinity College

\noindent Krista, L. D., Gallagher, P. T., \& Bloomfield, D. S. 2011, \apj, 731, 26

\noindent Landi, E. 2008, \apj, 685, 1270

\noindent Landi, E., \& Cranmer, S. R. 2009, \apj, 691, 794

\noindent Landi, E., Reale, F., \& Testa, P. 2012, \aap, 538, 10 

\noindent Lei, J., Thayer, J. P., Forbes, J. M., Sutton, E. K., \& Nerem, R. S. 2008, \grl, 35, L19105

\noindent Lilensten, J., Dudok de Wita, T., Amblard, P.-O., Aboudarham, J., Auchère, F., Kretzschmar, M., 
Annales Geophysicae, 2007, 25, 1299

\noindent Machiya, H. \& Akasofu, S.-I. 2014, Journal of Atmospheric and Solar-Terrestrial Physics, Volume 113, 44

\noindent Marsch, E., Tu, C.-Y., \& Wilhelm, K. 2000, \aap, 359, 381

\noindent McComas, D. J., Elliot, H. A \& von Steiger, R. 2002, Geophys.Res.Lette.,
29, 28

\noindent Mogilevsky, E. I., Obridko, V. N., \& Shilova, N. S. 1997, 176, 107

\noindent Moses, D., Clette, F., Delaboudini`ere, J-P. et al. 1997, Sol. Phys., 175, 571

\noindent Narukage, N., Sakao, T., Kano, R., Hara, H., Shimojo, M., Bando, T., Urayama, F., Deluca, E., Golub, L., Weber, M., Grigis, P., 

\noindent Navarro-Peralta, P. \& Sanchez-Ibarra, A. 1994, \solphys, 153
 Cirtain, J. \& Tsuneta, S. 2011, \solphys, 269, 169

\noindent Neupert, W.M. \& Pizzo, V. 1974, \jgr, 79, 3701

\noindent Nistico, G., Patsourakos, S., Bothmer, V., Zimbardo, G., Advances in Space Research, 48, 1490-

\noindent Nolte, J. T., Krieger, A. S., Timothy, A. F., Gold, R. E., Roelof, E. C., Vaiana, G., Lazarus, A. J., Sullivan, J. D., \&  McIntosh, P. S. 1976, \solphys 46, 303

\noindent Obridko, V. N. \& Shelting, B. D. 1989, \solphys, 124, 73

\noindent Obridko, V. N \& Solov’ev, A. A. 2011, Astronomy Reports,
55, 1144

\noindent Ofman, L. 2005, Space Science Reviews, 120, 67

\noindent Osherovich, V. A.; Gliner, E. B.; Tzur, I.; Kuhn, M. L. 1985, 
solar physics, 97, 251

\noindent Parker, E. N. 1955, \apj, 121, 491

\noindent Peter, H. \& Judge, P. G. 1999, \apj, 522, 1148

\noindent Ram, S. T., Liu, C. H., \& S.-Y. Su. 2010, \jgr, 115, 14

\noindent Roble, R. G. \& Schmidtke, G. 1979, Journal of Atmospheric and Terrestrial Physics, 41, 53

\noindent Richards, P. G., Fennelly, J. A., \& Torr, D. G. 1994, \jgr, 99, 8981

\noindent Rotter, T., Veronig, A. M., Temmer, M. \& Vršnak, B. 2012, \solphys, 281,793 

\noindent Shelke, R. N. \& Pande, M. C. 1985, \solphys, 95, 193

\noindent Shugai, Yu. S., Veselovsky, I. S., \& Trichtchenko, L. D. 2009, Ge\&Ae, 49, 415

\noindent Slemzin, V. A., Goryaev, F. F. \& Kuzin, S. V. 2014, Plasma Physics Reports, Volume 40, 855

\noindent Sojka, J. J., McPherron, R. L., van Eyken, A. P., Nicolls, M. J., Heinselman, C. J., \& Kelly, J. D. 2009, \grl, 36, L19105

\noindent Soon, W., Baliunas, S., Posmentier, E. S., \&  Okeke, P. 2000, NewAstronomy, 4, 563

\noindent Shugai, Y. S., Veselovsky, I. S \& Trichtchenko, L. D. 2009, 
Geomagnetism and Aeronomy, 49, Issue 4, pp.415-424

\noindent Stenflo, J. O. 1993, in Solar Surface Magnetism, ed. R. J. Rutten \& C. J. Shrijver
 (Dordrecht:Kluwer),8

\noindent Stucki, K., Solanki, S. K., Sch\"{u}hle, U., R\"{u}edi, I., Wilhelm, K., Stenflo, J. O., Brkovi\'{c}, A.
 \&  Huber, M. C. E. 2000, \aap, 363, 1145

\noindent Stucki, K., Solanki, S. K., Pike, C. D.,  Sch\"{u}hle, U., R\"{u}edi, I., Pauluhn, A., \& Brkovi\'{c}, A. 2002, \aap, 381, 653

\noindent Timothy, A. F. \& Krieger, A. S. 1975, \solphys, 42, 135

\noindent Tu, C.-Y., Marsch, E., Wilhelm, K. \& Curdt, W. 1998, \apj 503, 475

\noindent Verbanac, G., Vr\v{s}nak, B., Veronig, A., \& Temmer, M. 2011, \aap, 526, 20

\noindent Vršnak, B., Temmer, Ma Veronig, Astrid M. 2007,
\solphys, 240, 315

\noindent Wagner, W. J. 1975, \apj, 198L, 141

\noindent Wagner, W. J. 1976, Basic Mechanisms of Solar Activity,
Proceedings from IAU Symposium no. 71, edts. Bumba, V and Kleczek, J, p. 41

\noindent Wang, Y. M., \ssr, 2009, 144, 383

\noindent Warren, H. P., Mariska, J. T., Wilhelm, K., \apj, 1997, 490, L187

\noindent Webb, D. F. \&  Davis, J. M. 1985, \solphys, 102, 177

\noindent Wiegelmann, T., Thalmann, J, K. \& Solanki, S, K. 2014, The Astronomy and Astrophysics Review, 22, 106

\noindent Wilhelm, K., Marsch, E., Dwivedi, B. N., Hassler, D. M., Lemaire, P., Gabriel, A. H., \& Huber, M. C. E. 1998, \apj, 500, 1023

\noindent Wilhelm, K. 2006, \aap, 455, 697

\noindent Wilhelm, K. 2012, \ssr, 172, 57

\noindent Xia, L. D., Marsch, E. \& Wilhelm, K. 2004, \aap, 424, 1025

\noindent Yang, S.H., Zhang, J., Jin, C. L.,  Li, L. P  and H. Y. Duan, H. Y.
 2009, \aap, 501, 745
 
\noindent Zhang, J., White, S. M. \& Kundu, M. R. 1999, \apj, 527, 977

\noindent Zhang, H., Zhou, G., Wang, J. \&  Wang, H. 2007, \apj, 655, L113

\noindent Zhang, J., Zhou, G., Wang, J., \& Wang, H. 2007, \apj, 655, L113

\noindent Zirker, J. B. 1977, Reviews of Geophysics and Space Physics, 15, 257

\end{document}